\documentclass[12pt,preprint]{aastex}

\usepackage{natbib}

\def\deg{$^{\rm o}$}

\shortauthors{Gagne et al.}
\shorttitle{The Teacup AGN}

\begin{document}

\title{Spatially-Resolved Spectra of the ``Teacup'' AGN: Tracing the
History of a Dying Quasar}

\author{J.P. Gagne\altaffilmark{1},
D.M. Crenshaw\altaffilmark{1},
S.B. Kraemer\altaffilmark{2},
H.R. Schmitt\altaffilmark{3},
W.C. Keel\altaffilmark{4},
S. Rafter\altaffilmark{5},
T.C. Fischer\altaffilmark{1},
V.N. Bennert\altaffilmark{6},
K.Schawinski\altaffilmark{7}}

\altaffiltext{1}{Department of Physics and Astronomy, Georgia State 
University, Astronomy Offices, Twenty Five Park Place South SE, Suite 600,
Atlanta, GA 30303; gagne@chara.gsu.edu}

\altaffiltext{2}{Department of Physics, Catholic University of America, 620 Michigan Ave.,
N.E., Washington, DC 20064}

\altaffiltext{3}{Naval Research Laboratory, Washington, DC 20375}

\altaffiltext{4}{Department of Physics and Astronomy, University of
Alabama, Box 870324, Tuscaloosa, AL 35487, USA}

\altaffiltext{5}{Physics Department, Technion, Haifa 32000, Israel}

\altaffiltext{6}{Physics Department, California Polytechnic State
University San Luis Obispo, CA 93407}

\altaffiltext{7}{Institute for Astronomy, Department of Physics, ETH
Zurich, Wolfgang-Pauli-Strasse 27, CH-8093 Zurich, Switzerland}

\begin{abstract}
The Sloan Digital Sky Survey (SDSS) Galaxy Zoo project has revealed a
number of spectacular galaxies possessing Extended Emission-Line
Regions (EELRs), the most famous being Hanny's Voorwerp galaxy. We
present another EELR object discovered in the SDSS endeavor: the
Teacup Active Galactic Nucleus (AGN), nicknamed for its EELR, which
has a ``handle'' like structure protruding 15 kpc into the northeast
quadrant of the galaxy. We analyze physical conditions of this galaxy
with long-slit ground based spectroscopy from Lowell, Lick, and KPNO
observatories. With the Lowell 1.8-m Perkin’s telescope we took
multiple observations at different offset positions, allowing us to
recover spatially resolved spectra across the galaxy. Line diagnostics
indicate the ionized gas is photoionized primarily by the
AGN. Additionally we are able to derive the hydrogen density from the
[S~II] $\lambda$6716/$\lambda$6731 ratio. We generated two-component
photoionization models for each spatially resolved Lowell
spectrum. These models allow us to calculate the AGN bolometric
luminosity seen by the gas at different radii from the nuclear center
of the Teacup. Our results show a drop in bolometric luminosity by
more than two orders of magnitude from the EELR to the nucleus,
suggesting that the AGN has decreased in luminosity by this amount in
a continuous fashion over 46,000 years, supporting the case for a
dying AGN in this galaxy independent of any IR based evidence. We demonstrate
that spatially resolved photoionization modeling could be applied to
EELRs to investigate long time scale variability.
\end{abstract}


\section{Introduction}
\label{Introduction}

The Teacup AGN (SDSS J143029.88 +133912.0, 2MASX J14302986+1339117)
was originally discovered by Massimo Mezzoprete\footnote{\scriptsize
  http://www.galaxyzooforum.org/index.php?$\%$20topic=7539.msg62340}
in 2007 as part of the Sloan Digital Sky Survey Galaxy Zoo
project \citep{2012MNRAS.420..878K}. One of the campaigns during the
Galaxy Zoo project was to find AGN hosting Extended Emission-Line
Regions (EELRs), gaseous regions that are AGN-ionized at distances of
$\sim$10 kpc. Spectroscopically confirmed as an EELR by
\cite{2012MNRAS.420..878K}, the Teacup possesses a redshift of z =
0.086 to yield a distance of 344 Mpc (1$''$ $\sim$1.7 kpc for $H_0 = 73$ km
s$^{-1}$ Mpc$^{-1}$). The ``Teacup'' nickname was given to describe
its EELR, a spectacular "handle" or loop of ionized gas extending out
to $\sim$15 kpc northeast from the nucleus of the galaxy, a feature
best seen via HST imaging (Keel et al 2013; also posted
online\footnote{\scriptsize
  http://blog.galaxyzoo.org/2012/06/14/hubble-spies-the-teacup-and-i-spy-hubble/}). EELRs
offer a unique way to study AGN, by allowing us to study ionization
interactions between the central AGN engine and exterior regions of
the galaxy.

We use multiple sources of data available to us, the first being
images and spectra from the Sloan Digital Sky Survey. In Figure 1, we
show the SDSS image from the g, r and i bandpasses combined; in the
northeast corner the ionized handle is visible as a purple loop, which
can also be seen in the contour map. The continuum emission shows a
somewhat disturbed morphology, possibly from a merger or some other
interaction. Figure 2 shows a spectrum of the optical center of the
galaxy taken from SDSS with a spectral resolving power of R $\equiv$
$\lambda$/$\Delta\lambda$ $\approx$ 2000 and a 3$''$ diameter fiber;
strong emission lines characteristic of AGN-ionized gas are
present. Figure 3 presents one of our spectra taken from Lowell
Observatory at 5$''$ east from center, intersecting the loop
structured EELR. To explain the color of this feature in the SDSS
image, g, r and i filter response curves are overplotted. Note that
the g filter has been color coded to blue, r to green and i to
red. Interestingly, [O~III] $\lambda$5007 falls in between the g and r
bandpass filters. Because the rest of the emission lines fall inside
the ranges of both g and i filters while redshifted $\lambda$5007, the
strongest emission line, is clipped out by the bandpasses, the ionized
handle of gas in the NE appears purple in Figure 1, as does ionized
gas in the SW and other portions of the galaxy. The spectrum is
typical of a type 2 AGN (Khachikian \& Weedman 1971), with strong
emission lines covering a broad range in ionization, a weak continuum
with several stellar absorption features, and no evidence for broad
(FWHM $\geq$ 1000 km s$^{-1}$) emission lines.

In this paper, we present results from long-slit spectroscopic and
imaging observations from Lowell, KPNO, Lick, SDSS and VLA
observatories. The data provide an exciting opportunity to study the
long-term variability of AGN on time scales of $\leq$ 10$^{5}$
years. In $\S$\ref{Observations} we describe our methods for data
acquisition and reduction. In $\S$\ref{Analysis} we explain our
analysis of the data, yielding dereddend line ratios. Our
observational results are presented in $\S$\ref{Obs Results},
including kinematics of the Teacup AGN as well as FIRST radio data
from VLA. In $\S$\ref{Models} we describe our process in creating
photoionization models for the Teacup, and review the implications of
these models. In $\S$\ref{Discussion} we discuss our major findings
and conclusions.

\section{Observations and Data Reduction}
\label{Observations}

%We had multiple sources of data available to us, the first being
%images and spectra from the Sloan Digital Sky Survey. We show the SDSS
%image from g, r and i bandpasses combined in Figure 1; in the northeast
%corner the ionized handle is visible. The continuum emission shows a
%disturbed morphology, possibly from a merger or some other
%interaction. Figure 2 shows a spectrum of the optical center of the
%galaxy taken from SDSS with a spectral resolving power of R $\equiv$
%$\lambda$/$\Delta\lambda$ $\approx$ 2000 and a 3$''$ diameter
%fiber. The spectrum is typical of a type 2 AGN, with strong emission
%lines covering a broad range in ionization, a weak continuum with
%several stellar absorption features, and no evidence for broad (FWHM
%$\geq$ 1000 km s$^{-1}$) emission lines.

We obtained our primary spectroscopic observations on clear nights at Lowell
Observatory's Perkins 1.8-m reflector near Flagstaff, Arizona. We
obtained two dimensional long-slit spectra with the DeVeny ``blue''
spectrograph with a 300 line mm$^{-1}$ grating, a slit of 2$''$ width
positioned north-south, and a spectral resolution of $\Delta\lambda$
$\approx$ 3.0 \AA\,, with R ranging from 1400 - 2500, between 4300
$-$ 7600 \AA\,. A total of nine long-slit observations were obtained
at six parallel positions relative to the galactic center to cover the
12$''$$\times$12$''$ field containing the Teacup AGN. Four
observations were made at optical center; for these observations, we
averaged the line ratios, and the standard deviation between these
four data sets yielded our errors. Three observations were taken
moving eastward of the optical center at offset positions of 2$''$,
4$''$ and 5$''$ in order to obtain spectra across the Teacup's
ionized ``handle'', with the last two positions overlapping by
1$''$. Two more observations were made at 2$''$ and 4$''$ west of the
optical center. Because of differences in grating positioning in the
DeVeny spectrograph only data from slit positions 5$''$ east of
nucleus and nuclear center (0$''$) extend into the blue enough to
capture [O~II] $\lambda$3727.

  Additionally, we obtained long-slit spectra from the 2.1-m
  telescope under clear conditions at Kitt Peak National Observatory and 3.0-m Shane
  telescope at Lick Observatory. At KPNO we used the GoldCam
  spectrograph with the ``26new'' grating which is a 600 line mm$^{-1}$
  grating with $\Delta\lambda$ $\approx$ 3.3 \AA\, resolution. This
  2$''$ wide long-slit spectrometer was given a position angle of
  37\deg\, to capture the top side of the Teacup's ''handle''
  feature. Data collected from the LICK observatory was obtained with
  the Kast Double Spectrograph, which was set to a 2$''$ slit width
  and position angle of 95\deg\, covering the bottom side of the
  ionized ''handle''. To achieve the correct wavelength coverage we
  used the 600/7500 and 600/4310 grisms for the red and blue
  channels respectively.

Figure 4 shows the vertical 2$''$ slit positions from Lowell divided
into 31 2$''$$\times$2$''$ extraction bins. Overplotted are the KPNO
slit (position angle 37\deg\,) and the slit from Lick
Observatory (position angle of 95\deg\,); see also Keel et
al. 2012a. Lastly, data from the VLA FIRST survey
\citep{1995ApJ...450..559B} were analyzed in order to gain a radio
luminosity and classification of this object. Table 1 gives a full
account of all optical observations including that from SDSS. We
reduced our optical data two-dimensionally using
IRAF\footnote{\scriptsize IRAF is distributed by the National Optical
  Astronomy Observatories, which are operated by the Association of
  Universities for Research in Astronomy, Inc., under contract with
  the National Science Foundation.}, as well as the NOAO and CTIO
packages, resulting in wavelength calibrated spectral images in flux
units.

Two dimensional reduction allowed us to extract multiple spectra in
the cross-dispersion direction for each offset slit position in the
dispersion direction. While the width of the extraction bin
  is constrained to 2$''$ by the slit width,
  we had the ability to sample our data in the
  cross-dispersion direction in either 1$''$ or 2$''$
  bins. Extracting spectra in this manner from each of our six slit
  positions yielded a total of 62 or 31 individual spectra in
  2$''$$\times$1$''$ (3.4 kpc$\times$1.7 kpc) or
  2$''$$\times$2$''$ (3.4 kpc$\times$3.4 kpc) bins, respectively.
  Spectra sampled in 2$''$$\times$1$''$ bins have a higher spatial
  resolution but a lower signal-noise ratio than the data sampled in
  2$''$$\times$2$''$ bins. Therefore, our signal-noise requirements
  dictated when we could use data binned in 2$''$$\times$1$''$ or data
  in 2$''$$\times$2$''$ bins.

  To ensure the quality of our data, we required that the fluxes of
  the important emission lines for each type of analysis have
  signal-noise ratios $\ge$ 3$\sigma$. In the case of our kinematic
  analysis, the emission line of interest is [O~III] $\lambda$5007,
  because it is the strongest emission-line in our spectra. The
  strength of [O~III] $\lambda$5007 allowed us to use the
  2$''$$\times$1$''$ bins to obtain better spatial resolution for our
  kinematic measurements. For photoionization modeling, the lines of
  interest were the blended [S~II] $\lambda\lambda$ 6716, 6731
  doublet, to obtain accurate densities.  This feature is much weaker
  than the [O~III] $\lambda$5007, so we used 2$''$$\times$2$''$ bins.
  In the case of the 2$''$$\times$2$''$ bins, three low signal to
  noise bins from the 4$''$ west slit were discarded leaving 28 bins
  with acceptable signal to noise for the majority of our
  analysis. The seeing for the long slit spectroscopy was between 1"
  and 2" (FWHM of the point spread function), so some of the
  kinematics measurements could have been oversampled by up to a
  factor of two in the spatial direction.

  As the spectrum from SDSS
  has a larger aperture, greater wavelength coverage, and better
  signal to noise than the rest of our observations, we matched the
  absolute flux as a function of wavelength from SDSS to that of our
  central extracted bin from Lowell, in order to correct for
  atmospheric (including refraction) losses. We applied this
  normalization to the rest of the data for each night before analysis
  of emission line fluxes for each bin.

\section{Analysis}
\label{Analysis}

The series of bins allowed us to create a grid of spatially resolved
spectra covering the entire galaxy. We analyzed the reduced data
utilizing Interactive Data Language (IDL). For each spectral bin we
measured emission line fluxes for every non-blended line with a linear
interpolation procedure that integrates under each line and above a
baseline which connects the adjacent continua. For the blended lines
of [N~II] $\lambda$$\lambda$6548, 6583 and H$\alpha$, and the [S~II]
$\lambda$$\lambda$6716, 6731 doublet, we took a template from the
[O~III] $\lambda$5007 profile and reproduced it at the positions of
the blended components simultaneously. By scaling the template up and
down it was possible to find a scaling factor for each line relative
to [O~III]. To determine good fits between the templates and blended
lines, we subtracted the scaled components to look at the residuals,
rescaling until the residuals were close to the established continuum
ensuring good fits (Crenshaw \& Peterson 1986). All
continuum-subtracted fluxes were then divided by their corresponding
H$\beta$ fluxes in order to gain normalized spectral
ratios. Reddening, E(B$-$V), for each 2$''$$\times$2$''$ bin was found
by taking H$\alpha$/H$\beta$ = 2.9, appropriate for
recombination (Osterbrock \& Ferland 2006). To calculate the
  reddening across the entire spectral range for each emission line
  and correct each spectrum, we used a standard Galactic reddening
  curve described by \cite{1989ApJ...345..245C}. As shown in
  \cite{2001ApJ...555..633C}, reddening curves do not differ much
  in this region of the spectrum and any of a number of curves will
  provide a reasonable correction to observed NLR line ratios.

The values of all measured H$\beta$ fluxes and
E(B$-$V) values are listed by position in arcseconds relative to the
nucleus in Table 2. Extracted positions listed in each table are given
a Cartesian coordinate, where x corresponds to east (positive) and
west (negative) positions relative to the nuclear center, and y
corresponds to the north (positive) and south (negative)
directions. Following this convention, the nuclear center is labeled as
(0$''$,0$''$). The resultant normalized and dereddened line ratios can
be found in Tables 3 and 4. 

  We attribute the uncertainties in our flux measurements to a
  combination of photon noise (propagated through the data reduction
  process), measurement errors (including placement
  of the continuum), and errors in the reddening
  correction (propagated from the lines used to determine the reddening).
  To constrain our errors in measurement, we performed the
  same flux measurement procedure three times on each line with different reasonable
  continuum placements and
  found the standard deviation. The leading source of error in our
  reddening correction is from our errors associated
  with the measured H$\beta$ fluxes, as the H$\alpha$/H$\beta$ ratio
  was employed to determine E(B$-$V). These sources of error were all
  summed in quadrature, resulting in our final dereddened line ratio
  errors.

Dereddened normalized fluxes were employed in several diagnostic tests
using ``BPT'' diagrams as shown in Figure 5 (Baldwin et
al. 1981). According to these diagrams adopted from from
\cite{2006MNRAS.372..961K}, our emission line ratios indicate that the
Teacup AGN is indeed a Seyfert 2 undergoing photoionization from a
central ionizing source. There is no evidence from these diagrams for
a strong contribution from starbursts to the ionizing radiation.

 %The first plot compares
%log([O~III]/H$\beta$) vs
%log([N~II]/H$\alpha$) and is broken into two regions, an H II star
%formation region, and an AGN region. Our data lie in the AGN region
%of this graph. The second and third plots show the ratios of
%log([S~II]/H$\alpha$) and log([O~I]/H$\alpha$) respectively and serve
%to demarcate our galaxy further, between starburst galaxies, Low
%Ionization Narrow Emission-line Region galaxies (Heckman 1980) and
%Seyfert galaxies. 

Because of the relative strength of the [O~III] $\lambda$5007 emission
line we were able to employ our 2$''$$\times$1$''$ bins to map out
[O~III] flux and kinematics, as well as include the bins located at
4$''$ west. To probe large scale motions of the ionized gas in this
galaxy, kinematics were determined by fitting Gaussians to the [O~III]
$\lambda$5007 lines (see Fischer et al. 2010) and determining their
central wavelengths for all 62 bins spanning a 20.4 kpc $\times$ 28.9
kpc field. These wavelengths were converted to velocity offsets
from the rest frame of the galaxy, defined in this case by the
velocity of our centrally extracted bin. Our data do not have high
enough signal to noise to get a galactic velocity from stellar
absorption lines. The data from KPNO and Lick observatories provided
us with additional 2-D spectra which allowed us to investigate the
kinematics of the ionized loop of gas protruding from the nucleus of
the object. The FWHM([O~III]) was also measured for each Lowell bin;
the instrumental FWHM was subtracted off these values in quadrature to
obtain the intrinsic FWHM.

\section{Observational Results}
\label{Obs Results}
Having mapped out the Teacup with spatially resolved spectra, we were
able to investigate the emission line reddening, fluxes, and
kinematics in a grid over our object. Figure 6 shows a map of measured
E(B$-$V) reddening values per location on the left hand side. The
regions around (0$''$,2$''$) and ($-$2$''$,4$''$) show the highest
levels of reddening. The reddening map suggests a possible dust lane
running southeast to northwest at PA $\approx$ $-$45\deg\,. Figure 6
also shows [O~III] $\lambda$5007 flux for each 2$''$$\times$2$''$ bin,
measured in terms of ergs s$^{-1}$ cm$^{-2}$. The position
(0$''$,0$''$) shows the largest [O~III] flux of 9.7$\times$10$^{-14}$
ergs s$^{-1}$ cm$^{-2}$ at the location of the nucleus.

%fig 7 vel and fwhm maps
Our [O~III] velocity map, presented on the left of Figure 7, shows the
NE region of the ionized handle at position (2$''$,9$''$) receding
with a maximum redshifted velocity of 150 km s$^{-1}$ relative to the
center of the galaxy, while the position ($-$4$''$,$-$2$''$) has the
largest blueshifted velocity of 340 km s$^{-1}$. The velocity pattern
shows a clear kinematic axis running SE to NW with a low range in
radial velocities, indicative of rotational motion. Considering the
region sampled hosts a majority of the galaxy, the observed kinematics
must be due primarily to galactic rotation. This places the axis of
rotation at PA $\approx$ $-$50\deg\,. The right box in Figure 7 shows
measured FWHM values in terms of km s$^{-1}$, for the [O~III]
$\lambda$5007 emission line. In general, the FWHM increases toward the
central nucleus, which is also supporting evidence for galactic
rotation. However, the dust lane runs along the same direction as the
rotational axis, which is certainly not typical for normal disk
galaxies, suggesting disturbed kinematics, possibly due to a merger.

%fig 8 radial vel plots
Additional kinematic results from both KPNO 2.1-m and Lick Shane
3-m telescopes agree with this interpretation. As these observations
were made at different slit position angles, we created a procedure
that utilized our grid of Lowell observed velocities to simulate an
artificial slit at the KPNO/Lick observation position angles and then
extracted the pseudo-slit kinematics. Figure 8 displays radial
velocity curves from KPNO at 37\deg\ and LICK at 95\deg\ as well as an
overplot of our extracted velocity curves from Lowell. In general, the
velocities from the different data sets agree extremely well, and the
KPNO and Lick observations allow us to trace the radial velocities out
to greater distances. The radial velocity curve from the Lick data
turns over from a redshift to a blueshift at around 22$''$ northeast
of the nucleus. This suggests once again that there are other physical
phenomena affecting the kinematics at large distances in addition to
mere galactic rotation. This turnover is not seen in the velocity map
compiled from Lowell, because at 22$''$ away from the center, it is
outside the scope of our Lowell observations. A slight bump in the
radial velocity curves at the location of the teacup handle suggests
an additional kinematic component at this location, possibly outflow,
even though the magnitude is relatively low ($\sim$50 km s$^{-1}$).

%radio observations, fig 9 radio data
  We also investigate the radio properties of the Teacup. Figure
  9 shows a 20cm radio image from the FIRST survey. On the right of
  Figure 9 is the SDSS g-band image. The radio emission has a flux
  of 26.41$\pm$0.15 mJy (4.2$\times$$^{23}$~W~Hz$^{-1}$) and extends
  along the direction of the ”handle”. The radio image was deconvolved
  into two Gaussian components using the routine JMFIT in AIPS.  The
  stronger component, with a flux of 19.28$\pm$0.33 mJy corresponds to
  the nucleus. The second component, at 6.1′$''$ from the nucleus,
  along PA$=74^{\circ}$, has a flux of 5.86$\pm$0.15 mJy. There is a
  correspondence between the direction of the radio emission and the
  ionized gas in the handle. The radio luminosity of the Teacup is
  only 4.2$\times$10$^{23}$ watts Hz$^{-1}$, which is only slightly
  above the radio-loud dividing line of 1$\times$10$^{23}$ watts
  Hz$^{-1}$ (Best et al. 2005), making the Teacup a radio-intermediate
  AGN.

%Schmitt's old paragraph describing radio observations
%We also investigated the radio properties of the Teacup. Figure 9
%shows a 20cm radio image captured by FIRST which has been deconvolved
%into two Gaussians. On the right of Figure 9 is the SDSS g-band
%image \citep{1995ApJ...450..559B}. The radio emission has a flux of
%26.41$\pm$0.15 mJy (4.2$\times$$^{23}$~W~Hz$^{-1}$) and is extended
%along the direction of the "handle". We find the two decomposed
%Gaussians correspond to a stronger nuclear source with flux of
%19.28$\pm$0.33~mJy and a second component at 6.1$^{\prime\prime}$ from
%the nucleus, along PA$=74^{\circ}$, with a flux of 5.86$\pm$0.15
%mJy. There is a correspondence between the direction of the radio
%emission and the ionized gas in the handle. The radio luminosity of
%the Teacup is 4.2$\times$10$^{23}$ watts Hz$^{-1}$, which is only
%slightly above the radio-loud dividing line of 1$\times$10$^{23}$
%watts Hz$^{-1}$ (Best et al. 2005), making the Teacup a
%$''$radio-intermediate$''$ AGN.

\section{Photoionization Models}
\label{Models}
\subsection{\emph{Initial Conditions}}

To gain physical insight, we generated photoionization models for the
Teacup AGN with CLOUDY version 8.00 (see Ferland et al. 1998), which
replicates the observed line ratios by modeling a central ionizing
source acting upon a gaseous region, in this case modeled with a slab
geometry at each position (Kraemer et al. 1994). To satisfy signal to
noise requirements, we modeled each spectrum extracted from the 2$''$
by 2$''$ regions (excluding 4$''$ west data because of poor signal to
noise). In order to accurately model the photoionizing conditions of
the Teacup we first had to derive an expression for the spectral
energy distribution (SED). Our initial strategy followed from
\cite{1994ApJ...435..171K}, looking at the observed ratio of He II
$\lambda$4686 relative to H$\beta$. We determined the ionizing UV
power law index $\alpha$ from

\begin{equation}
R = \frac{Q(H^{0})}{Q(He^{+})} \approx \left(\frac{1}{4}\right)^{\alpha} - 1,
\end{equation}

\noindent
where R is the ratio of photons ($Q$) which are ionizing H$^{0}$ and
He$^{+}$. R can be determined from relative densities
($n$), intensities ($I$), and volume emissivities ($j$) of the two lines
in question, along with the dereddened ratio:

\begin{equation}
R = \frac{n(He)}{n(H)}\frac{I(H\beta)}{I(HeII\lambda4686)}\frac{j(HeII\lambda4686)}{j(H\beta)}.
\end{equation}

\noindent
Using this method for each extracted spectrum we arrived at an initial
average index of $\alpha$ = $-$1.7. However, upon inspection of our
initial models with this spectral index we found that we were
predicting 30$\%$ more He II/H$\beta$ than we were observing. For
this reason we decreased $\alpha$ to $-$2.0 which resulted in our
models matching He II/H$\beta$ to within $\pm$10 \%. We described our
final SED as a piecewise function, similar to that used by our group
for other AGN (Kraemer \& Crenshaw 2000):

\begin{equation}
\alpha = - 1.0,\hspace{12pt} \emph{h}\nu < 13.6 eV,
\end{equation} 

\begin{equation}
\alpha = - 2.0,\hspace{12pt} 13.6 eV \leq \emph{h}\nu < 1000 eV,
\end{equation}

\begin{equation}
\alpha = - 1.0,\hspace{12pt} \emph{h}\nu \geq 1000 eV,
\end{equation}

Previous work by \cite{2000ApJ...532..256K} has demonstrated solar
abundances are usually sufficient to accurately model the physical
conditions in Seyfert 2 narrow-line regions. Based on this experience
solar abundances were employed with and without dust depletion and we
found no appreciable difference. Because there is no direct evidence
for dust depletions or elemental peculiarities we assumed solar elemental
abundances \citep{2005ASPC..336...25A} without dust depletion. The
numerical values relative to Hydrogen for these abundances are: He =
0.10, C = 2.45$\times$10$^{-4}$, N = 6.03$\times$10$^{-5}$, O =
4.57$\times$10$^{-4}$, Ne = 6.92$\times$10$^{-5}$, Na =
2.09$\times$10$^{-6}$, Mg = 3.39$\times$10$^{-5}$, Al =
3.09$\times$10$^{-6}$, Si = 3.24$\times$10$^{-5}$, S =
1.38$\times$10$^{-5}$, Ar = 2.51$\times$10$^{-6}$, Ca =
2.24$\times$10$^{-6}$, Fe = 2.82$\times$10$^{-5}$, and Ni =
1.78$\times$10$^{-6}$.

By using the ratio of [S~II] $\lambda\lambda$
6716 to 6731 to estimate the density $n_{H}$ (Osterbrock \& Ferland
2006), our last free parameter is the unitless ionization parameter,
U, described by \cite{1983ApJ...264..105F}:

\begin{equation}
U = \frac{Q(H^{0})}{4\pi r^{2}n_{H}c},
\end{equation}

where $Q(H^{0})$ is the number of ionizing photons emitted by the
source per second, r is the distance from the central source to the
ionized cloud and $n_{H}$ is the hydrogen density of the gas. We
assumed that r is approximately the projected distance from each
extracted bin to the ionizing source, in this case simplified to be
bin (0$''$,0$''$).

\subsection{\emph{Model Components and Types}}

%single component modeling
Our initial models consisted of only a single component, with SED and
abundances set as described above. For each extracted position,
$n_{H}$ was set from the ratio of [S~II], leaving U and column depth
$N_{H}$ as free parameters. Values of U were then picked based on
specific line intensities such as [O~III] $\lambda$5007, [N~II]
$\lambda$6584 and [S~II] $\lambda$6731 given by
\cite{1983ApJ...264..105F}. Initial values for $N_{H}$ were chosen to
be $\sim$10$^{21}$ cm$^{-2}$, similar to values previously used by
Kraemer et al. (2000). Our initial values of U and $N_{H}$ were then
varied until our models fit our data to within a factor of two. Single
component models were sufficient to replicate most of the observed
emission lines. However these models were unable to match the observed
intensity of [O~II] $\lambda$3727 within an acceptable error defined
as the ratio of modeled line intensity divided by dereddened observed
line intensity, with a good fit having a ratio $\leq$ 2 and a perfect
fit with a ratio = 1. As our best single component models
underpredicted the [O~II] intensity by a factor of three or greater,
we included a second component into our CLOUDY models.

%double component modeling
Our two component models consisted of a low ionization component
(log(U) $\leq$ $-$3.0) and a high ionization component (log(U) $\geq$
$-$3.0). The low ionization component contributes the majority of the
emission lines including [O~II] $\lambda$3727, [O~I] $\lambda$6300,
[N~II] $\lambda\lambda$6548, 6584 and [S~II] $\lambda$$\lambda$ 6716,
6731, while the high ionization component contributes the majority of
the [O~III] $\lambda$5007 emission. Blending the line ratios of both
the high and low ionization components yields our composite model line
intensities, which we compared to our observed dereddened
values. Because our spectra are spatially resolved, we had to assure
that both high and low ionization components for each model were
co-located with the same distance r from the source. From equation 6,
we were able to constrain each co-located set with the condition
that log(U$_{low \ ion}$) + log($n_{H \ low \ ion}$) = log(U$_{high
  \ ion}$) + log($n_{H \ high \ ion}$). However, our two component
models still underpredicted [O~II] $\lambda$3727 by a factor of
$\sim$3 or more in most cases, forcing us to consider other possible
ionizing mechanisms, that might boost [O~II]. While our BPT diagrams
suggest this object is photoionized, we decided to explore the
possibilities of shock ionization.

%Allen's shock models
To consider shock ionization we matched both our observed FWHM values
and observed densities to the shock velocities and hydrogen densities
used in radiative shock models, generated by Allen et al. (2008),
comparing our data only to Allen's models with solar abundance with
the magnetic parameter ranging from B = 0.01$-$1.0 $\mu$G. Comparing
these results to those from photoionization, the photoionization
models still yielded an overall better fit to our observed spectra, as
the shock models underpredicted [O~II] $\lambda$3727 even more than
the photoionization models (factor of $\sim$11 versus a factor of
$\sim$3). Shock models fit [O~III] $\lambda$4363 marginally better,
with a factor of $\sim$2 too low, versus our CLOUDY models which were
a factor of $\sim$2.6 too low. For [O~III] $\lambda$5007 our average
photoionization models matched within a factor of $\sim$1 (almost a
perfect fit); the shock models matched with an average factor of
$\sim$1.5. Lastly for the [S~II] $\lambda$$\lambda$6716, 6731 lines,
we modeled with CLOUDY to an average factor of $\sim$1, while the
shock models were off by an average factor of $\sim$4. Considering the
mismatches associated with the shock models, we abandoned the idea of
shock ionization being the primary mode of ionization.

%double component models with cosmic rays Considering FIRST radio Next
We consider the possibility of heating by cosmic rays. Another way to
explain the excess [O~II] emission is radio plasma within the AGN,
which can act as a source of cosmic rays to heat the gas, as
previously shown by Ferland \& Mushotzky (1984) and
\cite{2009MNRAS.392.1475F}. In order to determine the density
  of cosmic rays in the handle region we used our FIRST radio
  measurements,described in the previous section.
The 20cm flux in the nuclear region corresponds to 20.5
mJy, while the flux in the handle is 5.9 mJy. Since the radio
observations were done with a resolution of $\sim$5$^{\prime\prime}$,
we are not able to use them to measure the size of the emitting
region. Instead, we use the Sloan $i$ band image (Figure 9), assuming
that the radio emission is confined inside the handle. We find that
the handle has a radius of 1.7$^{\prime\prime}$, which corresponds to
2.83$^{\prime\prime}$ if one assumes a Gaussian emitting region. Using
the minimum energy equations from Miley (1980) we can calculate the
magnetic field inside this region and the pressure and density of
relativistic particles ($n^{*}$). We assume a spectral index
$\alpha=0.7$, a ratio of energy in heavy particles to the energy in
electrons $k=1$ and a filling factor $\eta=1$. We find
B$_{min}=1.2\times10^{-5}$~G,
P$_{rel}=4.5\times10^{-12}$~dyne~cm$^{-2}$ and
n$_{e}=7.1\times10^{-9}$~cm$^{-3}$. These values have a small
dependence on the choice of $\alpha$, $k$ and $\eta$.  Using the scale
factor $\beta=10^5$ (Ferland \& Mushotzky 1984), which is defined as
the ratio of total heating to Coulomb heating, we calculate a density
of cosmic rays of $n^*=10^{-3}$~cm$^{-3}$. This value can be
considered an underestimate, considering the choice of $k$ and $\eta$
values, so we adopted log($n^{*}$)= $-$2.5~cm$^{-3}$.

Returning to our two component CLOUDY models, the expected cosmic ray
density of $n^{*}$ = 10$^{-2.5}$ cm$^{-3}$ was added to the parameters
in our low ionization component. Here we make the assumption that the
low ionization regions of gas have been compressed by radio plasma
jets, which results in the low ionization component having a higher
density than the high ionization component. This assumption also
implies that the low and high ionization components are displaced from
one another with respect to the radio plasma, although they are
assumed to be at roughly the same radial distances with respect to the
AGN. The resulting models were able to match the observed strong
intensities of [O~II] $\lambda$3727. However adding the cosmic rays
also added a collisional ionization factor to the gas, boosting the
intensities of [N~II] $\lambda$$\lambda$6548, 6584 and [S~II]
$\lambda$$\lambda$ 6716, 6731 to a factor of 2 or more higher than
observed. This unwanted increase in ionization due to collisions can
be moderated by applying a scaling law defined by \cite{hazy} which
describes how the cosmic ray density at the ionized face of the cloud
($n^{*}$) decreases as it propagates through the cloud. This is
described by a power law $n^{*}$ $\propto$ (r$^{*}$)$^{\alpha}$, where
r$^{*}$ is the depth of the cloud approximated by r$^{*}$
$\sim$10$\times$(N$_{H}$/n$_{H}$) and we started with power law
$\alpha$ = -2.5. We then varied values for r$^{*}$ and $\alpha$ until
proper fits were achieved.

%double component models with hextra
%We assumed that radio plasma compresses the higher ionization gas
%creating dense knots of lower ionization gas with an extra heating
%rate (Q) in terms of energy per volume per second (erg cm$^{-3}$
%s$^{-1}$), \cite{2007ApJ...668..730K}. To calculate this Q we
%considered the amount of heat cosmic rays added to the energy budget
%of our CLOUDY models. From that we were able to scale our extra heat
%input accordingly and add it into our low ionization component. The
%resulting models produced acceptable fits to our data, all within a
%factor of two or less from observed line intensities. While the values
%of [N~II] $\lambda$$\lambda$6548, 6584 and [S~II] $\lambda$$\lambda$
%6716, 6731 were still a little higher than observed, the extra heating
%models overall provided a better fit to our data than any other
%strategy we had previously tried. Tables 2 and 3 present all measured
%line ratios relative to H$\beta$ per position with modeled values for
%each line displayed underneath in parenthesis. The notation for
%position follows the same Cartesian scheme used throughout the paper
%thus far.

\subsection{\emph{Model Results}}
Our condition for an acceptable fit between our CLOUDY models and
observed spectra was to match the observed lines to within a factor of
two or less. This was achieved via two component CLOUDY models. The
observed strong intensity of [O~II] $\lambda$3727 proved to be a
difficult feature to replicate. However, by considering the addition
of cosmic rays as a result of radio plasma within the low ionization
gas we were able to increase the levels of [O~II] by a factor $>$ 3,
making it possible to match the observed [O~II] intensities as well as
the other lines. Modeled values are presented in Tables 3 $\&$ 4 in
parentheses under observed line intensities. Overall our modeled line
intensities fit our observed data within a factor of two or
less. However, the modeled values for [N~II] $\lambda\lambda$6548,
6583 and [S~II] $\lambda\lambda$6716, 6731 are often a factor of two
or stronger than what is observed. Our modeled values for strong
features such as [O~III] $\lambda$5007 are typically matched to within
90$\%$ of the observed intensity or better. Table 5 shows the position
for each spectrum, projected distance to the ionizing source (r), the
high ionization component values for ionization parameter (U),
hydrogen density ($n_{H}$), column depth ($N_{H}$), cloud depth
(r$^{*}$), as well as the low ionization parameter values for U,
$n_{H}$, $N_{H}$, r$^{*}$, cosmic ray density ($n^{*}$), cosmic ray
scaling law ($\alpha$), and r$^{*}$. The projected distance (r) of the
central position is a geometrical approximation as there is no
information on where the ionized gas is concentrated within the
central resolution element. At (0$''$,0$''$), r = 1.02 kpc,
which is the average distance from the edge of a 3.4 kpc$\times$3.4
kpc square to its center.

Errors associated with model input parameters U, $N_{H}$, r$^{*}$ and
$n^{*}$ were constrained iteratively by incrementally varying their
values after achieving a best fit model. Once a model no longer matched our
criteria for an acceptable fit, the difference between the incremented
parameter values and the best fit values yielded our errors. Errors for
$n_{H}$ came directly from our errors in measurement for [S~II]
$\lambda$$\lambda$ 6716, 6731. The uncertainty in projected distance
r is equal to the half width of each bin, as each model's location has
been approximated to the central position of each bin.

%The interaction of radio plasma could potentially produce
%microturbulence in the compressed pockets of low-ionization gas,
%\cite{2007ApJ...668..730K}. Kraemer et al. shows that the velocity of
%microturbulence ($v_{turb}$) is a function of the heating rate (Q),
%hydrogen density ($n_{H}$) and column density ($N_{H}$). Therefore,
%for each spatially resolved model we were able to calculate $v_{turb}$
%in km s$^{-1}$ for each extracted spectrum. This is a velocity
%dispersion, so we converted to FWHM by multiplying by 2.355 assuming a
%Gaussian profile. Table 5 shows all calculated values of $v_{turb}$
%per position. Comparing the calculated values of FWHM($v_{turb}$) for
%each position to the measured FWHM values per position (Figure 6.,
%Table 5), we find the FHWM values on average are $\sim$ 200 km
%s$^{-1}$ greater than the values of FWHM($v_{turb}$). Thus, it is
%possible that microturbulence is present, but there must be an
%additional broadening agent such as macroturbulence present as well.

Once we had acceptable models for each spatially resolved spectrum, we
were able to determine $Q(H^{0})$ for each 2$''$$\times$2$''$ bin from
equation 6. By integrating our piecewise SED for energies greater than
1 eV, we can derive the ratio of $Q(H^{0})$ and bolometric luminosity
($L_{Bol}$). The ratio of $Q(H^{0})$/$L_{Bol}$ = 6.5$\times$10$^{9}$
photons erg$^{-2}$, allowed us to convert from our calculated
$Q(H^{0})$ to $L_{Bol}$ seen by the gas at each position. Table 5
shows the log of the derived bolometric luminosity ($L_{Bol}$) for
each modeled position. The errors in $L_{Bol}$ are propagated
  uncertainties in
  ionization parameter U, radial distance r, and hydrogen
  density ($n_{H}$). For the central region (0$''$,0$''$), r is by
  far the leading source of error in $L_{Bol}$ due to the large
  fractional uncertainty in r for this position. The contribution of
  uncertainty in $L_{Bol}$ decreases quickly for regions at
  greater distances of r.

Figure 10 gives a visual representation of
$L_{Bol}$ for each position, which shows that the lowest bolometric
luminosities reside at the nuclear center and surrounding regions,
while the regions further out and close to the handle seem to be
experiencing much higher levels of luminosity. We found the lowest
bolometric luminosity at the nuclear center (0$''$,0$''$) to be
2.3$\times$10$^{44}$ erg s$^{-1}$, more than two orders of magnitude
less luminous than that of region (2$''$,8$''$), the furthest region
sampled from the source, at which $L_{Bol}$ = 2.2$\times$10$^{46}$ erg
s$^{-2}$, $\sim$14.1 kpc away from its ionizing source. This drop in
luminosity, by a factor greater than 90 over a time frame of
$\sim$4$\times$10$^{4}$ years, fits nicely within the time frame of
0.2$-$2$\times$10$^{5}$ years reported by \cite{2012MNRAS.420..878K}
for AGN possessing EELRs such as Hanny's Voorwerp/IC 2497.

Figure 11 plots projected distance from the nucleus in kpc versus the
log of our bolometric luminosities. Note, due to the way our data were
sampled, we have multiple extracted bins that share the same radial
distance from the nuclear center. Therefore, luminosities at the same
radial distance have been offset by $\pm$0.1$^{''}$ in position to see
our error bars. There is no significant differences between east,
west, and center points compared to fit. Luminosity increasing with
radius as shown in Figures 10 and 11 suggest that the luminosity of
the Teacup's central AGN is decreasing dramatically with time.

Although $L_{Bol}$ is a function of U, r and $n_{H}$, it is
  worth pointing out that small deviations of these
  variables within our observational constraints will still yield the
  same result of $L_{Bol}$ increasing with radial distance away from
  the Teacup's nucleus. We find no evidence from our observed spectra
  that ionization parameter drops off from the central AGN
  within the observed region. Furthermore,
  the addition of cosmic rays to our models has no effect on the
  values used for ionization parameter. Thus adding cosmic rays does
  not change the increase in $L_{Bol}$ with radius as calculated from
  our models.

\section{Discussion and Conclusions}
\label{Discussion}

We have obtained long-slit spectra of the Teacup AGN in the optical
range of 3500$-$9000 \AA\, resulting in spatially resolved spectra. Our kinematic
results from the measured [O~III] $\lambda$5007 centroids suggest
primarily galactic rotation, with the major axis of rotation at PA
$\approx$ $-$50\deg\,. The area in the NE corner of the galaxy where
the EELR is located is red-shifted, while the SW area is blue-shifted,
relative to the galactic nucleus. However the morphology of the
galaxy, the dust lane along the rotation axis, and the turnover of the
radial velocity curves suggest a more complex situation, possibly due
to a merger. Radial velocity plots from KPNO and Lick spectra show a
slight bump in velocity over the area of the Teacup's EELR, suggesting
there could be something more than galactic rotation acting on the
ionized handle. One possibility could be that the EELR is a slow
moving outflow. The magnitude of observed velocities is quite low
compared to what is normally observed in nearby AGN outflows,
(Crenshaw et al. 2000), but this would be an outflow on much larger
scales, up to 15 kpc from the nucleus.

From our analysis of BPT diagrams we were able to diagnose the Teacup
as a Type 2 AGN undergoing photoionization. Using CLOUDY we created
photoionization models for a grid of 28 spatially resolved spectra
gathered from Lowell observatory. Each spectrum was replicated with
composite models consisting of high and low ionization components. Our
models match our observed emission lines quite well (within a factor
of two or less). To achieve acceptable fits to the [O~II] emission, we
had to include cosmic rays from the radio plasma in our models. The
cosmic ray density was approximated using the radio luminosity of this
object.

From our CLOUDY models we were able to calculate values of
bolometric luminosity for each position, and find that $L_{Bol}$
increases radially from the nuclear center (0$''$,0$''$). We interpret
this finding as strong evidence that the central AGN engine decreased
in luminosity by a factor $\sim$90 over a period $\sim$46,000 years.
\cite{2012MNRAS.420..878K} demonstrated luminosity fading on
timescales between 0.2$-$2$\times$10$^{5}$ years for multiple Galaxy
Zoo objects with EELRs, including Hanny's Voorwerp and the Teacup AGN
This conclusion was based on the strong ionizing flux needed to create
the EELR, the lack of a bright AGN, and the lack of strong IR flux that
would indicate an obscured AGN. Keel et al. concluded that the nuclear
IR luminosities are
  not enough to produce the levels of ionization observed in these
  object's EELRs, indicating the AGN output has decreased with
  time. \cite{2012AJ....144...66K} then bolstered this claim with a
  detailed study of the EELR Hanny's Voorwerp and its host galaxy
  IC2497. Using Space Telescope Imaging Spectrograph (STIS) data they
  found that Hanny's Voorwerp exhibits higher levels of ionization in
  the optical regime than the central AGN of IC2497. Using a
  combination of X-ray and IR data, Keel et al. are able to derive a
  spectral energy distribution for the AGN. Once again they find that
  the AGN within IC2497 does not possess enough energy to account for
  observed levels of ionization in Hanny's
  Voorwerp. \cite{2012AJ....144...66K} show that the ionizing
  luminosity of IC2497's AGN has dropped by a factor of $>$ 100 in the
  last (1-2)$\times$10$^{5}$ years.

Because we have measured
luminosities at multiple distances from the ionizing nucleus, we can
extend the analysis to study changes in luminosity over radius and thus time. Our
results show a continuous drop in luminosity for the Teacup AGN within
a timescale consistent with previous results by Keel et
al. Thus, we have additional evidence showing results that agree with previous
findings that are independent of IR luminosity arguments.

The Teacup AGN presents two additional interesting characteristics:
One is the disturbed morphological loop of the northeast region. The
second interesting phenomenon is its extremely strong [O~II]
$\lambda$3727 emission. One possible explanation for both is that the
loop was created by expanding radio-jet plasma, consistent with the
direction of the current jet, and cosmic rays from the radio plasma
also boost the [O~II] intensities to what we observe. The galaxy 3C 48
shows both similar morphology and [O~II] $\lambda$3727/H$\beta$ ratios
within its narrow-line region \citep{2007ApJ...659..195S}. Further
studies of these objects might offer new perspectives on the
interaction between radio-jet plasma and ionized gas in AGN.

The EELR around the ionized handle experienced the greatest
luminosity, with the location (2$''$,8$''$) being greater than 90
times more luminous than the center at $\sim$46,000 light years
away. Considering the change in luminosity across the distance between
the nucleus and the NE corner at (2$''$,8$''$), the time frame for the
drop in luminosity fits in with previous studies (Keel et
al. 2012). From our results, the Teacup's nucleus is dimming at
$\sim$4.7$\times$10$^{41}$ erg s$^{-1}$ per year, so that 46,000 years
ago it was luminous enough to qualify it as a type 2 quasar, while
currently the nucleus is in the luminosity regime of a Seyfert 2. From
our data presented in Figure 11 we see luminosity is decreasing
approximately exponentially. At the current rate at which the Teacup
is dimming, it will fall out of the realm of Seyfert luminosity at
$\sim$10$^{43}$ erg s$^{-1}$, in another $\sim$470 years. To further
probe the underlying physics behind this change in luminosity, as well
the spectacular ionized handle of gas extending out in the
north-east corner, further data should be collected. Analysis of
high resolution imaging from HST will yield a better look at the
extended structure of the handle. X-ray observations would allow a
probe of the central AGN, and radio observations would provide the
opportunity to look for possible jets, which could be driving the
handle.

%%%%%%%%%%%%%%%%%%%%%%ACKNOWLEDGEMENTS%%%%%%%%%%%%%%%%%%%%%%%
KS gratefully acknowledges support from Swiss National Science
Foundation Grant PP00P2\_138979/1. JG would like to acknowledge KS for
the naming of the Teacup AGN, as well as thank DMC and SBK for their
mentorship.

%%%%%%%%%%%%%%%%%%%%%%REFERENCES%%%%%%%%%%%%%%%%%%%%%%%%%%%%

\clearpage
%%%%%%%%%%%%%%%%%%%%%%TABLES%%%%%%%%%%%%%%%%%%

\clearpage

\begin{deluxetable}{lccccccc}
\rotate
\tablecolumns{5}
\footnotesize
\tablecaption{Observation list for the Teacup AGN}

\tablewidth{0pt}
\centering
\tablehead{
\colhead{Date} & \colhead{Observatory} & \colhead{Telescope} & \colhead{P.A.} & \colhead{Slit Position} & \colhead{Exposure Time} & \colhead{Wavelength}     \\
               &                       &                     &                & \colhead{from Nucleus}  & \colhead{(s)}           & \colhead{Coverage (\AA)} \\ 
                                 
}
\startdata
2005 May 13    & SDSS     & Survey 2.5-m   & $0$\deg   & Center     & 2880 & 3800-9250       \\
2009 April 18  & Lowell   & Perkins 1.8-m  & $0$\deg   & Center     & 1200 & 3860-7650       \\
2009 April 18  & Lowell   & Perkins 1.8-m  & $0$\deg   & 5$''$ East & 1200 & 3860-7650       \\
2010 May 20    & Lowell	   & Perkins 1.8-m  & $0$\deg   & Center     & 1200 & 3870-7660       \\
2010 June 15   & KPNO     & 2.1-m          & $37$\deg  & Center     & 2700 & 3270-5710       \\
2010 July 16   & LICK     & Shane 3.0-m    & $95$\deg  & Center     & 1800 & 3500-5600       \\
2010 July 16   & LICK     & Shane 3.0-m    & $95$\deg  & Center     & 1800 & 5540-8340       \\   
2011 May 3     & Lowell   & Perkins 1.8-m  & $0$\deg   & Center     & 1800 & 3930-7730       \\
2011 May 3     & Lowell   & Perkins 1.8-m  & $0$\deg   & 2$''$ West & 1800 & 3930-7730       \\ 
2011 May 3     & Lowell   & Perkins 1.8-m  & $0$\deg   & 4$''$ West & 1800 & 3930-7730       \\   
2011 May 5     & Lowell   & Perkins 1.8-m  & $0$\deg   & Center     & 1800 & 3930-7730       \\
2011 May 5     & Lowell   & Perkins 1.8-m  & $0$\deg   & 2$''$ East & 1800 & 3930-7730       \\
2011 May 5     & Lowell   & Perkins 1.8-m  & $0$\deg   & 4$''$ East & 1800 & 3930-7730       \\
\enddata

\end{deluxetable}

\clearpage

\scriptsize
\centering
\begin{deluxetable}{lcccc}
%\rotate
%\raggedleft
\footnotesize
\tablecaption{E(B-V) and H$\beta$ flux by position}
\tablewidth{0pt}
%\voffset{000pt}
\setlength{\tabcolsep}{0.09in}
%\label{tab:phot}
\tabletypesize{\scriptsize}
\tablehead{

\colhead{Position}             &\colhead{E(B-V)}    &\colhead{H$\beta$ flux}                         \\
\colhead{($x^{''}$,$y^{''}$)}  &\colhead{(mag)}     &\colhead{(ergs s$^{-1}$ cm$^{-2}$)}             \\
}
\startdata
($5,4$)                   & 0.17 $\pm$ 0.14     & 1.21$\times$10$^{-15}$ ($\pm$ 0.21)                          \\
($5,2$)                   & 0.00 $\pm$ 0.37     & 1.78$\times$10$^{-15}$ ($\pm$ 0.23)                          \\
($5,0$)                   & 0.00 $\pm$ 0.31     & 1.54$\times$10$^{-15}$ ($\pm$ 0.42)                          \\
($5,-2$)                  & 0.11 $\pm$ 0.47     & 0.57$\times$10$^{-15}$ ($\pm$ 0.16)                          \\
($4,6$)                   & 0.16 $\pm$ 0.14     & 0.67$\times$10$^{-15}$ ($\pm$ 0.09)                          \\
($4,4$)                   & 0.21 $\pm$ 0.16     & 1.40$\times$10$^{-15}$ ($\pm$ 0.19)                          \\
($4,2$)                   & 0.21 $\pm$ 0.19     & 1.63$\times$10$^{-15}$ ($\pm$ 0.16)                          \\
($4,0$)                   & 0.40 $\pm$ 0.15     & 1.50$\times$10$^{-15}$ ($\pm$ 0.16)                          \\
($4,-2$)                  & 0.35 $\pm$ 0.24     & 0.90$\times$10$^{-15}$ ($\pm$ 0.16)                          \\
($4,-4$)                  & 0.27 $\pm$ 0.18     & 0.38$\times$10$^{-15}$ ($\pm$ 0.13)                          \\
($2,8$)                   & 0.14 $\pm$ 0.19     & 0.36$\times$10$^{-15}$ ($\pm$ 0.12)                          \\
($2,6$)                   & 0.03 $\pm$ 0.18     & 1.15$\times$10$^{-15}$ ($\pm$ 0.18)                          \\
($2,4$)                   & 0.22 $\pm$ 0.14     & 2.47$\times$10$^{-15}$ ($\pm$ 0.30)                          \\
($2,2$)                   & 0.35 $\pm$ 0.10     & 4.40$\times$10$^{-15}$ ($\pm$ 0.44)                          \\
($2,0$)                   & 0.53 $\pm$ 0.08     & 3.97$\times$10$^{-15}$ ($\pm$ 0.30)                          \\
($2,-2$)                  & 0.60 $\pm$ 0.15     & 2.03$\times$10$^{-15}$ ($\pm$ 0.31)                          \\
($2,-4$)                  & 0.43 $\pm$ 0.29     & 0.76$\times$10$^{-15}$ ($\pm$ 0.18)                          \\
($0,6$)                   & 0.04 $\pm$ 0.29     & 1.11$\times$10$^{-15}$ ($\pm$ 0.33)                          \\
($0,4$)                   & 0.43 $\pm$ 0.20     & 2.08$\times$10$^{-15}$ ($\pm$ 1.21)                          \\
($0,2$)                   & 0.62 $\pm$ 0.11     & 6.82$\times$10$^{-15}$ ($\pm$ 1.84)                          \\
($0,0$)                   & 0.53 $\pm$ 0.10     & 12.72$\times$10$^{-15}$ ($\pm$ 3.02)                         \\
($0,-2$)                  & 0.41 $\pm$ 0.13     & 4.76$\times$10$^{-15}$ ($\pm$ 0.44)                          \\
($0,-4$)                  & 0.26 $\pm$ 0.20     & 0.90$\times$10$^{-15}$ ($\pm$ 0.27)                          \\
($-2,4$)                  & 0.70 $\pm$ 0.17     & 1.21$\times$10$^{-15}$ ($\pm$ 0.21)                          \\
($-2,2$)                  & 0.40 $\pm$ 0.09     & 6.08$\times$10$^{-15}$ ($\pm$ 0.62)                          \\
($-2,0$)                  & 0.26 $\pm$ 0.07     & 7.91$\times$10$^{-15}$ ($\pm$ 0.52)                          \\
($-2,-2$)                 & 0.43 $\pm$ 0.15     & 2.94$\times$10$^{-15}$ ($\pm$ 0.43)                          \\
($-2,-4$)                 & 0.66 $\pm$ 0.22     & 0.57$\times$10$^{-15}$ ($\pm$ 0.11)                          \\

\enddata
\normalsize
\end{deluxetable}   

\clearpage

\scriptsize
\centering
\begin{deluxetable}{lcccccccccc}
\rotate
%\raggedleft
\footnotesize
\tablecaption{Dereddend Line Ratios ($\lambda 3727$ - $\lambda 4686$)\tablenotemark{a}\,\,\tablenotemark{b}}

\tablewidth{0pt}
%\voffset{000pt}
\setlength{\tabcolsep}{0.09in}
%\label{tab:phot}
\tabletypesize{\scriptsize}
\tablehead{

\colhead{Position}             & \colhead{[O~II]}         & \colhead{$[$Ne~III$]$}   & \colhead{H$8$,He I}      & \colhead{H$\epsilon$,[Ne~III]}      & \colhead{H$\delta$ $\lambda 4100$} & \colhead{H$\gamma$}      & \colhead{$[$O~III$]$}    & \colhead{He II}           \\
\colhead{($x^{''}$,$y^{''}$)}   & \colhead{$\lambda 3727$} & \colhead{$\lambda 3869$} & \colhead{$\lambda 3889$} & \colhead{$\lambda 3970$}            & \colhead{$\lambda 4100$}           & \colhead{$\lambda 4341$} & \colhead{$\lambda 4363$} & \colhead{$\lambda 4686$}  \\
}

\startdata
($5,4$)                    & $10.07 \pm 1.25$         & $1.19 \pm 0.13$          & --                       & $0.45 \pm 0.04$                     & --                                 & $0.60 \pm 0.04$          & --                       & --                        \\
                           & ($7.28$)                   & ($0.76$)                   & ($0.19$)                   & ($0.39$)                              & ($0.26$)                             & ($0.47$)                   & ($0.07$)                   & ($0.42$)                    \\          
($5,2$)                    & $8.00 \pm 1.48$          & $0.96 \pm 0.16$          & $0.25 \pm 0.04$          & $0.34 \pm 0.05$                     & $0.19 \pm 0.02$                    & $0.41 \pm 0.04$          & $0.14\pm 0.01$           & $0.25 \pm 0.01$           \\
                           & ($7.28$)                   & ($0.76$)                   & ($0.19$)                   & ($0.39$)                              & ($0.26$)                             & ($0.47$)                   & ($0.07$)                   & ($0.42$)                    \\
($5,0$)                    & $8.17 \pm 2.23$          & $0.88 \pm 0.21$          & $0.20 \pm 0.05$          & $0.31 \pm 0.07$                     & --                                 & $0.38 \pm 0.05$          & --                       & --                        \\
                           & ($7.37$)                   & ($0.65$)                   & ($0.21$)                   & ($0.35$)                              & ($0.26$)                             & ($0.47$)                   & ($0.05$)                   & ($0.14$)                    \\
($5,-2$)                   & $9.99 \pm 4.18$          & $1.41 \pm 0.51$          & --                       & $0.93 \pm 0.30$                     & --                                 & $0.90 \pm 0.18$          & --                       & --                        \\
                           & ($7.08$)                   & ($0.70$)                   & ($0.19$)                   & ($0.37$)                              & ($0.26$)                             & ($0.47$)                   & ($0.06$)                   & ($0.34$)                    \\
($4,6$)                    & --                       & $1.08 \pm 0.11$          & --                       & $0.51 \pm 0.05$                     & $0.23 \pm 0.02$                    & $0.58 \pm 0.03$          & --                       & --                        \\
                           & ($7.31$)                   & ($0.76$)                   & ($0.19$)                   & ($0.39$)                              & ($0.26$)                             & ($0.47$)                   & ($0.07$)                   & ($0.42$)                    \\
($4,4$)                    & --                       & $0.98 \pm 0.12$          & --                       & $0.47 \pm 0.05$                     & $0.38 \pm 0.04$                    & $0.37 \pm 0.03$          & $0.10 \pm 0.01$          & --                        \\
                           & ($6.06$)                   & ($0.83$)                   & ($0.20$)                   & ($0.41$)                              & ($0.26$)                             & ($0.47$)                   & ($0.10$)                   & ($0.14$)                    \\
($4,2$)                    & --                       & $1.01 \pm 0.15$          & --                       & $0.50 \pm 0.07$                     & $0.20 \pm 0.02$                    & $0.65 \pm 0.05$          & $0.21 \pm 0.02$          & --                        \\
                           & ($5.11$)                   & ($0.68$)                   & ($0.20$)                   & ($0.36$)                              & ($0.26$)                             & ($0.47$)                   & ($0.06$)                   & ($0.19$)                    \\
($4,0$)                    & --                       & $1.07 \pm 0.12$          & --                       & $0.54 \pm 0.06$                     & $0.20 \pm 0.02$                    & $0.45 \pm 0.03$          & $0.15 \pm 0.01$          & $0.16 \pm 0.003$          \\
                           & ($5.11$)                   & ($0.68$)                   & ($0.20$)                   & ($0.36$)                              & ($0.26$)                             & ($0.47$)                   & ($0.06$)                   & ($0.19$)                    \\
($4,-2$)                   & --                       & $1.12 \pm 0.20$          & --                       & --                                  & --                                 & $0.49 \pm 0.05$          & --                       & --                        \\
                           & ($4.25$)                   & ($0.60$)                   & ($0.18$)                   & ($0.34$)                              & ($0.26$)                             & ($0.47$)                   & ($0.06$)                   & ($0.48$)                    \\
($4,-4$)                   & --                       & $0.45 \pm 0.06$          & --                       & --                                  & --                                 & --                       & --                       & --                        \\
                           & ($4.30$)                   & ($0.52$)                   & ($0.21$)                   & ($0.32$)                              & ($0.26$)                             & ($0.47$)                   & ($0.03$)                   & ($0.14$)                    \\
($2,8$)                    & --                       & $1.30 \pm 0.18$          & --                       & --                                  & --                                 & --                       & --                       & --                        \\
                           & ($4.25$)                   & ($0.61$)                   & ($0.18$)                   & ($0.35$)                              & ($0.26$)                             & ($0.47$)                   & ($0.06$)                   & ($0.47$)                    \\
($2,6$)                    & --                       & $0.88 \pm 0.12$          & --                       & $0.38 \pm 0.05$                     & --                                 & $0.56 \pm 0.04$          & --                       & $0.12 \pm 0.003$          \\
                           & ($4.31$)                   & ($0.61$)                   & ($0.20$)                   & ($0.34$)                              & ($0.26$)                             & ($0.47$)                   & ($0.05$)                   & ($0.31$)                    \\
($2,4$)                    & --                       & $0.91 \pm 0.10$          & $0.27 \pm 0.03$          & $0.30 \pm 0.03$                     & $0.25 \pm 0.02$                    & $0.42 \pm 0.03$          & $0.11 \pm 0.01$          & $0.11 \pm 0.002$          \\
                           & ($7.36$)                   & ($0.69$)                   & ($0.20$)                   & ($0.37$)                              & ($0.26$)                             & ($0.47$)                   & ($0.04$)                   & ($0.22$)                    \\
($2,2$)                    & --                       & $0.79 \pm 0.06$          & --                       & $0.23 \pm 0.02$                     & $0.20 \pm 0.01$                    & $0.44 \pm 0.02$          & $0.10 \pm 0.004$         & $0.12 \pm 0.002$          \\
                           & ($10.06$)                  & ($0.82$)                   & ($0.15$)                   & ($0.39$)                              & ($0.24$)                             & ($0.46$)                   & ($0.18$)                   & ($0.15$)                    \\
($2,0$)                    & --                       & $0.85 \pm 0.05$          & --                       & $0.22 \pm 0.01$                     & $0.25 \pm 0.01$                    & $0.45 \pm 0.02$          & $0.19 \pm 0.01$          & $0.16 \pm 0.002$          \\
                           & ($5.11$)                   & ($0.68$)                   & ($0.20$)                   & ($0.36$)                              & ($0.26$)                             & ($0.47$)                   & ($0.06$)                   & ($0.19$)                    \\
($2,-2$)                   & --                       & $1.25 \pm 0.14$          & --                       & $0.14 \pm 0.01$                     & $0.14 \pm 0.01$                    & --                       & --                       & $0.09 \pm 0.002$          \\
                           & ($5.11$)                   & ($0.68$)                   & ($0.20$)                   & ($0.36$)                              & ($0.26$)                             & ($0.47$)                   & ($0.06$)                   & ($0.19$)                    \\
($2,-4$)                   & --                       & $0.64 \pm 0.14$          & --                       & --                                  & --                                 & --                       & --                       & --                        \\
                           & ($8.70$)                   & ($0.75$)                   & ($0.13$)                   & ($0.37$)                              & ($0.24$)                             & ($0.46$)                   & ($0.14$)                   & ($0.38$)                    \\
($0,6$)                    & --                       & $0.79 \pm 0.18$          & --                       & $0.42 \pm 0.09$                     & --                                 & $0.39 \pm 0.05$          & --                       & --                        \\
                           & ($10.44$)                  & ($0.72$)                   & ($0.13$)                   & ($0.36$)                              & ($0.24$)                             & ($0.47$)                   & ($0.10$)                   & ($0.41$)                    \\
($0,4$)                    & $11.02 \pm 1.86$         & $1.15 \pm 0.17$          & --                       & $0.37 \pm 0.05$                     & $0.33 \pm 0.04$                    & $0.51 \pm 0.04$          & $0.17 \pm 0.01$          & $0.15 \pm 0.004$          \\
                           & ($10.15$)                  & ($0.85$)                   & ($0.13$)                   & ($0.40$)                              & ($0.26$)                             & ($0.46$)                   & ($0.16$)                   & ($0.38$)                    \\
($0,2$)                    & $9.31 \pm 0.83$          & $1.22 \pm 0.10$          & $0.24 \pm 0.02$          & $0.34 \pm 0.02$                     & $0.22 \pm 0.01$                    & $0.63 \pm 0.03$          & $0.19 \pm 0.01$          & --                        \\
                           & ($7.04$)                   & ($0.94$)                   & ($0.19$)                   & ($0.45$)                              & ($0.26$)                             & ($0.47$)                   & ($0.12$)                   & ($0.42$)                    \\
($0,0$)                    & $6.22 \pm 0.52$          & $0.91 \pm 0.07$          & $0.27 \pm 0.02$          & $0.24 \pm 0.02$                     & $0.18 \pm 0.01$                    & $0.48 \pm 0.02$          & $0.12 \pm 0.01$          & --                        \\
                           & ($6.17$)                   & ($0.83$)                   & ($0.19$)                   & ($0.41$)                              & ($0.26$)                             & ($0.47$)                   & ($0.09$)                   & ($0.04$)                    \\
($0,-2$)                   & $8.01 \pm 0.91$          & $1.07 \pm 0.11$          & $0.22 \pm 0.02$          & $0.25 \pm 0.02$                     & $0.21 \pm 0.02$                    & $0.55 \pm 0.03$          & $0.17 \pm 0.01$          & --                        \\
                           & ($8.70$)                   & ($0.75$)                   & ($0.13$)                   & ($0.37$)                              & ($0.24$)                             & ($0.46$)                   & ($0.14$)                   & ($0.38$)                    \\
($0,-4$)                   & $11.74 \pm 2.20$         & $1.43 \pm 0.23$          & $0.23 \pm 0.04$          & --                                  & --                                 & $0.60 \pm 0.06$          & --                       & --                        \\
                           & ($8.70$)                   & ($0.75$)                   & ($0.13$)                   & ($0.37$)                              & ($0.24$)                             & ($0.46$)                   & ($0.14$)                   & ($0.38$)                    \\
($-2,4$)                   & --                       & $0.85 \pm 0.11$          & --                       & --                                  & --                                 & --                       & --                       & --                        \\
                           & ($4.26$)                   & ($0.63$)                   & ($0.19$)                   & ($0.35$)                              & ($0.26$)                             & ($0.47$)                   & ($0.06$)                   & ($0.37$)                    \\
($-2,2$)                   & --                       & $0.85 \pm 0.06$          & --                       & $0.17 \pm 0.01$                     & $0.11 \pm 0.01$                    & $0.45 \pm 0.02$          & --                       & --                        \\
                           & ($6.03$)                   & ($0.82$)                   & ($0.19$)                   & ($0.41$)                              & ($0.26$)                             & ($0.47$)                   & ($0.09$)                   & ($0.37$)                    \\
($-2,0$)                   & --                       & $0.72 \pm 0.04$          & --                       & $0.23 \pm 0.01$                     & $0.12 \pm 0.005$                   & $0.40 \pm 0.01$          & --                       & --                        \\
                           & ($10.08$)                  & ($0.78$)                   & ($0.13$)                   & ($0.38$)                              & ($0.24$)                             & ($0.47$)                   & ($0.13$)                   & ($0.43$)                    \\
($-2,-2$)                  & --                       & $0.76 \pm 0.09$          & --                       & $0.27 \pm 0.03$                     & --                                 & $0.41 \pm 0.03$          & --                       & --                        \\
                           & ($10.13$)                  & ($0.72$)                   & ($0.13$)                   & ($0.36$)                              & ($0.24$)                             & ($0.46$)                   & ($0.09$)                   & ($0.38$)                    \\
($-2,-4$)                  & --                       & $1.79 \pm 0.30$          & --                       & --                                  & --                                 & --                       & --                       & --                        \\
                           & ($8.37$)                   & ($0.80$)                   & ($0.12$)                   & ($0.38$)                              & ($0.24$)                             & ($0.47$)                   & ($0.18$)                   & ($0.43$)                    \\

\enddata
%\normalsize
\tablenotetext{a}{Model values are given in parentheses}
\tablenotetext{b}{Positions presented in Cartesian format (x,y), with positive x value for east of center and negative for west.
By the same format positive y denotes north of center and negative for south. The center position is marked by (0,0).}
\end{deluxetable}

\clearpage

\scriptsize
\centering
\begin{deluxetable}{lccccccccc}
\rotate
%\raggedleft
\footnotesize
\tablecaption{Dereddend Line Ratios($\lambda4959$ - $\lambda6731$)\tablenotemark{a}}

\tablewidth{0pt}
%\voffset{000pt}
\setlength{\tabcolsep}{0.09in}
%\label{tab:phot}
\tabletypesize{\scriptsize}
\tablehead{

\colhead{Position}            & \colhead{$[$O~III$]$}    & \colhead{$[$O~III$]$}    & \colhead{He I}           & \colhead{$[$O~I$]$}      & \colhead{$[$O~I$]$}      & \colhead{$[$N~II$]$}     &  \colhead{$[$N~II$]$}     & \colhead{$[$S~II$]$}     & \colhead{$[$S~II$]$}      \\
\colhead{($x^{''}$,$y^{''}$)} & \colhead{$\lambda 4959$} & \colhead{$\lambda 5007$} & \colhead{$\lambda 5876$} & \colhead{$\lambda 6300$} & \colhead{$\lambda 6364$} & \colhead{$\lambda 6548$} &  \colhead{$\lambda 6583$} & \colhead{$\lambda 6716$} & \colhead{$\lambda 6731$}  \\
}

\startdata

($5,4$)                  & $2.64 \pm 0.03$          & $8.33 \pm 0.13$          & --                       & $0.36 \pm 0.05$          & --                       & $0.35 \pm 0.05$          & $1.20 \pm 0.19$           & $0.49 \pm 0.08$          & $0.56 \pm 0.09$           \\
                       & ($2.55$)                   & ($7.68$)                   & ($0.09$)                   & ($0.21$)                   & ($0.07$)                   & ($0.55$)                   & ($1.63$)                    & ($0.89$)                   & ($0.88$)                    \\  
($5,2$)                  & $2.13 \pm 0.03$          & $7.10 \pm 0.16$          & --                       & $0.40 \pm 0.08$          & --                       & $0.30 \pm 0.07$          & $1.12 \pm 0.27$           & $0.52 \pm 0.13$          & $0.59 \pm 0.15$           \\
                       & ($2.55$)                   & ($7.68$)                   & ($0.09$)                   & ($0.21$)                   & ($0.07$)                   & ($0.55$)                   & ($1.63$)                    & ($0.89$)                   & ($0.88$)                    \\
($5,0$)                  & $2.09 \pm 0.04$          & $6.77 \pm 0.22$          & --                       & $0.29 \pm 0.09$          & --                       & $0.41 \pm 0.14$          & $1.15 \pm 0.40$           & $0.68 \pm 0.25$          & $0.54 \pm 0.20$           \\
                       & ($2.03$)                   & ($6.11$)                   & ($0.13$)                   & ($0.21$)                   & ($0.07$)                   & ($0.55$)                   & ($1.62$)                    & ($0.93$)                   & ($0.87$)                    \\   
($5,-2$)                 & $2.17 \pm 0.07$          & $7.17 \pm 0.36$          & --                       & $0.59 \pm 0.27$          & --                       & $0.27 \pm 0.14$          & $0.97 \pm 0.53$           & --                       & --                        \\ 
                       & ($2.24$)                   & ($6.75$)                   & ($0.10$)                   & ($0.58$)                   & ($0.18$)                   & ($0.57$)                   & ($1.67$)                    & ($1.19$)                   & ($1.11$)                    \\
($4,6$)                  & $2.35 \pm 0.02$          & $8.16 \pm 0.12$          & --                       & --                       & --                       & $0.29 \pm 0.04$          & $1.13 \pm 0.17$           & $0.63 \pm 0.10$          & $0.63 \pm 0.10$           \\
                       & ($2.64$)                   & ($7.93$)                   & ($0.09$)                   & ($0.08$)                   & ($0.03$)                   & ($0.53$)                   & ($1.56$)                    & ($0.76$)                   & ($0.77$)                    \\
($4,4$)                  & $2.55 \pm 0.03$          & $7.73 \pm 0.13$          & --                       & $0.44 \pm 0.07$          & --                       & $0.29 \pm 0.05$          & $1.13 \pm 0.20$           & $0.69 \pm 0.13$          & $0.61 \pm 0.12$           \\
                       & ($3.11$)                   & ($9.36$)                   & ($0.12$)                   & ($0.49$)                   & ($0.16$)                   & ($0.43$)                   & ($1.27$)                    & ($0.86$)                   & ($0.79$)                    \\
($4,2$)                  & $2.67 \pm 0.03$          & $7.78 \pm 0.16$          & --                       & $0.50 \pm 0.09$          & --                       & $0.39 \pm 0.08$          & $1.07 \pm 0.23$           & $0.60 \pm 0.14$          & $0.60 \pm 0.14$           \\
                       & ($2.59$)                   & ($7.80$)                   & ($0.12$)                   & ($0.36$)                   & ($0.12$)                   & ($0.43$)                   & ($1.26$)                    & ($0.76$)                   & ($0.76$)                    \\
($4,0$)                  & $2.72 \pm 0.03$          & $7.53 \pm 0.12$          & --                       & $0.38 \pm 0.05$          & --                       & $0.34 \pm 0.06$          & $0.96 \pm 0.16$           & $0.59 \pm 0.11$          & $0.62 \pm 0.11$           \\
                       & ($2.59$)                   & ($7.80$)                   & ($0.12$)                   & ($0.36$)                   & ($0.12$)                   & ($0.43$)                   & ($1.26$)                    & ($0.76$)                   & ($0.76$)                    \\
($4,-2$)                 & $2.01 \pm 0.03$          & $6.53 \pm 0.17$          & --                       & $0.43 \pm 0.10$          & --                       & $0.35 \pm 0.09$          & $1.14 \pm 0.31$           & $0.54 \pm 0.15$          & $0.70 \pm 0.20$           \\
                       & ($2.16$)                   & ($6.51$)                   & ($0.09$)                   & ($0.29$)                   & ($0.09$)                   & ($0.45$)                   & ($1.33$)                    & ($0.65$)                   & ($0.78$)                    \\
($4,-4$)                 & $1.27 \pm 0.02$          & $5.40 \pm 0.10$          & --                       & --                       & --                       & $0.35 \pm 0.07$          & $1.15 \pm 0.22$           & --                       & --                        \\
                       & ($1.83$)                   & ($5.50$)                   & ($0.13$)                   & ($0.29$)                   & ($0.09$)                   & ($0.45$)                   & ($1.34$)                    & ($0.65$)                   & ($0.79$)                    \\
($2,8$)                  & $1.76 \pm 0.02$          & $6.19 \pm 0.12$          & --                       & --                       & --                       & $0.33 \pm 0.07$          & $1.45 \pm 0.30$           & $0.69 \pm 0.15$          & $0.85 \pm 0.19$           \\
                       & ($2.24$)                   & ($6.73$)                   & ($0.09$)                   & ($0.29$)                   & ($0.09$)                   & ($0.45$)                   & ($1.33$)                    & ($0.65$)                   & ($0.78$)                    \\
($2,6$)                  & $2.03 \pm 0.02$          & $6.64 \pm 0.13$          & --                       & $0.28 \pm 0.05$          & --                       & $0.39 \pm 0.08$          & $1.16 \pm 0.23$           & $0.65 \pm 0.14$          & $0.77 \pm 0.16$           \\
                       & ($2.22$)                   & ($6.67$)                   & ($0.11$)                   & ($0.47$)                   & ($0.15$)                   & ($0.47$)                   & ($1.38$)                    & ($0.80$)                   & ($0.90$)                    \\
($2,4$)                  & $2.35 \pm 0.02$          & $7.46 \pm 0.11$          & $0.10 \pm 0.01$          & $0.41 \pm 0.06$          & $0.16 \pm 0.02$          & $0.44 \pm 0.07$          & $1.20 \pm 0.19$           & $0.64 \pm 0.11$          & $0.59 \pm 0.10$           \\
                       & ($2.26$)                   & ($6.82$)                   & ($0.12$)                   & ($0.21$)                   & ($0.07$)                   & ($0.55$)                   & ($1.62$)                    & ($0.93$)                   & ($0.87$)                    \\ 
($2,2$)                  & $2.33 \pm 0.02$          & $7.19 \pm 0.08$          & $0.15 \pm 0.01$          & $0.42 \pm 0.04$          & $0.08 \pm 0.01$          & $0.39 \pm 0.04$          & $1.22 \pm 0.14$           & $0.59 \pm 0.07$          & $0.49 \pm 0.06$           \\
                       & ($2.49$)                   & ($7.50$)                   & ($0.07$)                   & ($0.41$)                   & ($0.13$)                   & ($0.41$)                   & ($1.21$)                    & ($1.03$)                   & ($0.82$)                    \\
($2,0$)                  & $2.47 \pm 0.01$          & $7.80 \pm 0.07$          & $0.10 \pm 0.01$          & $0.35 \pm 0.03$          & $0.09 \pm 0.01$          & $0.35 \pm 0.03$          & $1.19 \pm 0.11$           & $0.48 \pm 0.05$          & $0.44 \pm 0.04$           \\
                       & ($2.59$)                   & ($7.80$)                   & ($0.12$)                   & ($0.36$)                   & ($0.12$)                   & ($0.43$)                   & ($1.26$)                    & ($0.76$)                   & ($0.76$)                    \\
($2,-2$)                 & $2.41 \pm 0.02$          & $7.81 \pm 0.13$          & --                       & $0.63 \pm 0.05$          & --                       & $0.35 \pm 0.06$          & $1.16 \pm 0.20$           & $0.54 \pm 0.10$          & $0.50 \pm 0.09$           \\
                       & ($2.59$)                   & ($7.80$)                   & ($0.12$)                   & ($0.36$)                   & ($0.12$)                   & ($0.43$)                   & ($1.26$)                    & ($0.76$)                   & ($0.76$)                    \\
($2,-4$)                 & $2.35 \pm 0.05$          & $8.24 \pm 0.26$          & --                       & $0.60 \pm 0.17$          & --                       & $0.49 \pm 0.16$          & $1.45 \pm 0.48$           & $0.68 \pm 0.24$          & $0.54 \pm 0.19$           \\
                       & ($2.50$)                   & ($7.54$)                   & ($0.04$)                   & ($0.51$)                   & ($0.16$)                   & ($0.35$)                   & ($1.03$)                    & ($0.98$)                   & ($0.78$)                    \\
($0,6$)                  & $1.77 \pm 0.04$          & $5.47 \pm 0.17$          & --                       & $0.35 \pm 0.10$          & --                       & $0.42 \pm 0.14$          & $1.16 \pm 0.38$           & $0.68 \pm 0.24$          & $0.55 \pm 0.19$           \\
                       & ($2.07$)                   & ($6.23$)                   & ($0.04$)                   & ($0.40$)                   & ($0.13$)                   & ($0.41$)                   & ($1.22$)                    & ($0.98$)                   & ($0.84$)                    \\
($0,4$)                  & $2.17 \pm 0.03$          & $7.40 \pm 0.15$          & $0.09 \pm 0.01$          & $0.39 \pm 0.07$          & --                       & $0.42 \pm 0.09$          & $1.34 \pm 0.29$           & $0.56 \pm 0.13$          & $0.53 \pm 0.12$           \\
                       & ($2.71$)                   & ($8.16$)                   & ($0.04$)                   & ($0.42$)                   & ($0.14$)                   & ($0.39$)                   & ($1.15$)                    & ($0.99$)                   & ($0.79$)                    \\
($0,2$)                  & $2.70 \pm 0.02$          & $8.34 \pm 0.09$          & $0.10 \pm 0.01$          & $0.45 \pm 0.04$          & --                       & $0.46 \pm 0.05$          & $1.33 \pm 0.15$           & $0.42 \pm 0.05$          & $0.39 \pm 0.05$           \\
                       & ($3.45$)                   & ($10.39$)                  & ($0.09$)                   & ($0.28$)                   & ($0.09$)                   & ($0.49$)                   & ($1.45$)                    & ($0.79$)                   & ($0.81$)                    \\
($0,0$)                  & $2.56 \pm 0.02$          & $7.64 \pm 0.08$          & $0.11 \pm 0.01$          & $0.34 \pm 0.03$          & --                       & $0.44 \pm 0.05$          & $1.29 \pm 0.14$           & $0.43 \pm 0.05$          & $0.39 \pm 0.04$           \\
                       & ($3.16$)                   & ($9.50$)                   & ($0.10$)                   & ($0.28$)                   & ($0.09$)                   & ($0.46$)                   & ($1.35$)                    & ($0.79$)                   & ($0.77$)                    \\
($0,-2$)                 & $2.66 \pm 0.02$          & $8.29 \pm 0.12$          & $0.12 \pm 0.01$          & $0.45 \pm 0.06$          & --                       & $0.45 \pm 0.06$          & $1.31 \pm 0.19$           & $0.52 \pm 0.08$          & $0.43 \pm 0.07$           \\
                       & ($2.50$)                   & ($7.54$)                   & ($0.04$)                   & ($0.51$)                   & ($0.16$)                   & ($0.35$)                   & ($1.03$)                    & ($0.98$)                   & ($0.78$)                    \\
($0,-4$)                 & $2.58 \pm 0.04$          & $7.71 \pm 0.18$          & --                       & $0.49 \pm 0.10$          & --                       & $0.47 \pm 0.11$          & $1.33 \pm 0.32$           & $0.54 \pm 0.14$          & $0.42 \pm 0.11$           \\ 
                       & ($2.50$)                   & ($7.71$)                   & ($0.04$)                   & ($0.51$)                   & ($0.16$)                   & ($0.35$)                   & ($1.03$)                    & ($0.98$)                   & ($0.78$)                    \\
($-2,4$)                 & $1.94 \pm 0.02$          & $7.26 \pm 0.13$          & --                       & $0.62 \pm 0.10$          & --                       & $0.48 \pm 0.09$          & $1.19 \pm 0.22$           & $0.37 \pm 0.07$          & $0.40 \pm 0.08$           \\
                       & ($2.42$)                   & ($7.29$)                   & ($0.10$)                   & ($0.29$)                   & ($0.09$)                   & ($0.45$)                   & ($1.33$)                    & ($0.64$)                   & ($0.78$)                    \\
($-2,2$)                 & $2.98 \pm 0.02$          & $9.13 \pm 0.09$          & --                       & $0.47 \pm 0.04$          & --                       & $0.47 \pm 0.05$          & $1.41 \pm 0.15$           & $0.45 \pm 0.05$          & $0.48 \pm 0.05$           \\
                       & ($3.11$)                   & ($9.35$)                   & ($0.10$)                   & ($0.38$)                   & ($0.12$)                   & ($0.46$)                   & ($1.34$)                    & ($0.85$)                   & ($0.80$)                    \\
($-2,0$)                 & $2.47 \pm 0.01$          & $7.66 \pm 0.06$          & --                       & $0.38 \pm 0.03$          & --                       & $0.48 \pm 0.04$          & $1.42 \pm 0.11$           & $0.62 \pm 0.05$          & $0.44 \pm 0.04$           \\
                       & ($2.34$)                   & ($7.05$)                   & ($0.04$)                   & ($0.44$)                   & ($0.14$)                   & ($0.41$)                   & ($1.21$)                    & ($0.98$)                   & ($0.83$)                    \\
($-2,-2$)                & $1.86 \pm 0.02$          & $6.00 \pm 0.10$          & --                       & $0.46 \pm 0.07$          & --                       & $0.36 \pm 0.06$          & $1.21 \pm 0.21$           & $0.64 \pm 0.12$          & $0.36 \pm 0.07$           \\
                       & ($2.12$)                   & ($6.39$)                   & ($0.04$)                   & ($0.47$)                   & ($0.15$)                   & ($0.43$)                   & ($1.25$)                    & ($1.03$)                   & ($0.88$)                    \\
($-2,-4$)                & $2.18 \pm 0.03$          & $7.73 \pm 0.18$          & --                       & $0.55 \pm 0.12$          & --                       & $0.37 \pm 0.09$          & $1.04 \pm 0.26$           & $1.00 \pm 0.26$          & $0.42 \pm 0.11$           \\
                       & ($2.76$)                   & ($8.31$)                   & ($0.03$)                   & ($0.50$)                   & ($0.16$)                   & ($0.34$)                   & ($0.99$)                    & ($0.94$)                   & ($0.75.$)                    \\

\enddata
%\normalsize
\tablenotetext{a}{Model values are given in parentheses}
\end{deluxetable}

\clearpage

\scriptsize
\centering
\begin{deluxetable}{lccccccccccc}
\rotate
%\raggedleft
\footnotesize
\tablecaption{Projected distance r as well as U, $n_{H}$ and N$_{H}$ for both model components, $n^{*}$, $\alpha$ and r$^{*}$ for low ionization component, and L$_{Bol}$\tablenotemark{a}}

\tablewidth{0pt}
%\voffset{000pt}
\setlength{\tabcolsep}{0.09in}
%\label{tab:phot}
\tabletypesize{\scriptsize}
\tablehead{

\colhead{Position}              &\colhead{r}        &\colhead{High Ion}  &\colhead{High Ion}        &\colhead{High Ion}                  &\colhead{Low Ion}     &\colhead{Low Ion}                &\colhead{Low Ion}             &\colhead{Low Ion}       &\colhead{Low Ion}             &\colhead{Low Ion}                         &\colhead{log($L_{Bol}$)}   \\    
\colhead{($x^{''}$,$y^{''}$)}    &\colhead{(kpc)}    &\colhead{log(U)}    &\colhead{log($n_{H}$)}    &\colhead{log(N$_{H}$)}               &\colhead{log(U)}      &\colhead{log($n_{H}$)}            &\colhead{log(N$_{H}$)}        &\colhead{log($n^{*}$)}   &\colhead{$\alpha$}          &\colhead{log(r$^{*}$)}    &\colhead{(erg s$^{-1}$)}     \\               
\colhead{}                      &\colhead{}         &\colhead{}          &\colhead{(cm$^{-3}$)}     &\colhead{(cm$^{-2}$)}                &\colhead{}            &\colhead{(cm$^{-3}$)}              &\colhead{(cm$^{-2}$)}        &\colhead{(cm$^{-3}$)}    &\colhead{}                  &\colhead{(cm)}                             &\colhead{}                   \\
}																											       																			  
\startdata																										       																		  
																											       																			  
($5,4$)                           & 10.88           & -2.4               & 1.8$^{+0.50}_{-0.10}$                     & 19.6                  & -3.4              & 2.8$^{+0.50}_{-0.10}$                    & 19.7                   & -2.5                  & -3.5               & 18.2                  & 46.22$^{+0.16}_{-0.14}$  \\
($5,2$)                           &  9.18           & -2.4	        & 1.8$^{+1.10}_{-0.50}$			  & 19.6           	   & -3.4		& 2.8$^{+1.10}_{-0.50}$      	         & 19.7                   & -2.5  	          & -3.5               & 18.2                   & 46.07$^{+0.24}_{-0.18}$ \\
($5,0$)                           &  8.50           & -2.3	        & 1.6$^{+0.80}_{-1.70}$			  & 20.5		   & -3.4		& 2.7$^{+0.80}_{-1.70}$      	         & 19.7                   & -2.5  	          & -4.0               & 17.7                   & 45.90$^{+0.22}_{-0.32}$ \\
($5,-2$)                          &  9.18           & -2.4	        & 1.7$^{+0.50}_{-0.90}$			  & 19.8		   & -3.5		& 2.8$^{+0.50}_{-0.90}$      	         & 19.7                   & -2.5  	          & -4.0               & 17.7                   & 45.97$^{+0.18}_{-0.21}$ \\
($4,6$)                           & 12.24           & -2.4	        & 1.8$^{+0.30}_{-0.50}$			  & 19.6		   & -3.4		& 2.8$^{+0.30}_{-0.50}$      	         & 19.6                   & -2.5  	          & -3.5               & 17.7                   & 46.32$^{+0.13}_{-0.15}$ \\
($4,4$)                           &  9.69           & -2.4	        & 2.1$^{+0.30}_{-1.80}$			  & 20.4		   & -3.1		& 2.8$^{+0.30}_{-1.80}$      	         & 20.2                   & -2.5  	          & -3.5               & 18.5                   & 46.42$^{+0.16}_{-0.32}$ \\
($4,2$)                           &  7.65           & -2.4	        & 2.1$^{+0.30}_{-0.60}$			  & 20.2		   & -3.2		& 2.9$^{+0.30}_{-0.60}$      	         & 20.1                   & -2.5  	          & -5.0               & 18.2                   & 46.21$^{+0.20}_{-0.21}$ \\
($4,0$)                           &  6.80           & -2.4	        & 2.1$^{+0.30}_{-0.60}$			  & 20.2		   & -3.2		& 2.9$^{+0.30}_{-0.60}$      	         & 20.1                   & -2.5  	          & -5.0               & 18.2                   & 46.11$^{+0.22}_{-0.24}$ \\
($4,-2$)                          &  7.65           & -2.3	        & 2.1$^{+0.90}_{-0.30}$			  & 20.5		   & -3.4		& 3.2$^{+0.90}_{-0.30}$      	         & 19.8                   & -2.4        	  & -3.5               & 17.6                   & 46.31$^{+0.23}_{-0.20}$ \\
($4,-4$)                          &  9.69           & -2.3	        & 2.1$^{+0.50}_{-0.90}$			  & 20.5		   & -3.4		& 3.2$^{+0.50}_{-0.90}$      	         & 19.8                   & -2.5                  & -3.5               & 17.6                   & 46.52$^{+0.17}_{-0.20}$ \\
($2,8$)                           & 14.11           & -2.3	        & 1.6$^{+0.70}_{-0.30}$			  & 20.5		   & -3.4		& 2.7$^{+0.70}_{-0.30}$      	         & 19.7                   & -2.5        	  & -4.0               & 17.7                   & 46.34$^{+0.16}_{-0.12}$ \\
($2,6$)                           & 10.20           & -2.3	        & 1.6$^{+0.50}_{-0.30}$			  & 20.5		   & -3.4		& 2.7$^{+0.50}_{-0.30}$      	         & 19.7                   & -2.5  	          & -4.0               & 17.7                   & 46.10$^{+0.16}_{-0.15}$ \\
($2,4$)                           &  7.65           & -2.4	        & 1.8$^{+0.40}_{-0.80}$			  & 19.6		   & -3.4		& 2.8$^{+0.40}_{-0.80}$      	         & 19.7                   & -2.5  	          & -3.5               & 18.2                   & 45.91$^{+0.20}_{-0.23}$ \\
($2,2$)                           &  4.76           & -2.4	        & 2.1$^{+0.50}_{-1.40}$			  & 20.2		   & -3.2		& 2.9$^{+0.50}_{-1.40}$      	         & 20.1                   & -2.5  	          & -5.0               & 18.2                   & 45.80$^{+0.32}_{-0.38}$ \\
($2,0$)                           &  3.40           & -2.4	        & 2.1$^{+0.10}_{-0.70}$			  & 20.2		   & -3.2        	& 2.9$^{+0.10}_{-0.70}$      	         & 20.1                   & -2.5  	          & -5.0               & 18.2                   & 45.51$^{+0.44}_{-0.45}$ \\
($2,-2$)                          &  4.76           & -2.4	        & 2.1$^{+0.20}_{-0.80}$			  & 20.2		   & -3.2		& 2.9$^{+0.20}_{-0.80}$      	         & 20.1                   & -2.5  	          & -5.0               & 18.2                   & 45.80$^{+0.31}_{-0.33}$ \\
($2,-4$)                          &  7.65           & -2.4	        & 2.1$^{+1.40}_{-1.50}$			  & 20.4		   & -3.1		& 2.8$^{+1.40}_{-1.50}$      	         & 20.2                   & -2.5  	          & -3.5               & 18.5                   & 46.21$^{+0.29}_{-0.30}$ \\
($0,6$)                           & 10.20           & -2.3	        & 1.6$^{+1.00}_{-1.60}$			  & 20.5		   & -3.4		& 2.7$^{+1.00}_{-1.60}$      	         & 19.7                   & -2.5  	          & -4.0               & 17.7                   & 46.06$^{+0.22}_{-0.30}$ \\
($0,4$)                           &  6.80           & -2.4	        & 1.7$^{+0.80}_{-1.40}$			  & 20.0		   & -3.4		& 2.7$^{+0.80}_{-1.40}$      	         & 19.7                   & -2.5  	          & -3.5               & 18.0                   & 45.71$^{+0.25}_{-0.31}$ \\
($0,2$)                           &  3.40           & -2.4	        & 1.7$^{+0.10}_{-0.80}$			  & 19.6		   & -3.4		& 2.7$^{+0.10}_{-0.80}$      	         & 19.7                   & -2.5  	          & -3.5               & 18.0                   & 45.11$^{+0.44}_{-0.45}$ \\
($0,0$)                           &  1.02           & -2.4	        & 1.7$^{+0.10}_{-0.70}$			  & 19.6		   & -3.2		& 2.8$^{+0.10}_{-0.70}$      	         & 20.0                   & -2.5  	          & -2.5               & 18.4                   & 44.36$^{+1.45}_{-1.45}$ \\
($0,-2$)                          &  3.40           & -2.4	        & 1.7$^{+0.50}_{-1.40}$			  & 19.6		   & -3.4		& 2.7$^{+0.50}_{-1.40}$      	         & 19.8                   & -2.6  	          & -2.5               & 18.0                   & 45.11$^{+0.44}_{-0.49}$ \\
($0,-4$)                          &  6.80           & -2.4	        & 1.7$^{+0.70}_{-1.40}$			  & 19.6		   & -3.4		& 2.7$^{+0.70}_{-1.40}$      	         & 19.7                   & -2.5  	          & -3.5               & 18.0                   & 45.71$^{+0.25}_{-0.31}$ \\
($-2,4$)                          &  7.65           & -2.4	        & 2.1$^{+0.40}_{-0.80}$			  & 20.2		   & -3.2		& 2.9$^{+0.40}_{-0.80}$      	         & 20.1                   & -2.5        	  & -5.0               & 18.2                   & 46.21$^{+0.20}_{-0.22}$ \\
($-2,2$)                          &  4.76           & -2.4	        & 2.1$^{+0.30}_{-0.10}$			  & 19.6		   & -3.2		& 2.9$^{+0.30}_{-0.10}$      	         & 20.0                   & -2.4  	          & -2.0               & 18.1                   & 45.40$^{+0.31}_{-0.31}$ \\
($-2,0$)                          &  3.40           & -2.4	        & 2.1$^{+0.50}_{-0.50}$			  & 20.2		   & -3.2		& 2.9$^{+0.50}_{-0.50}$      	         & 20.1                   & -2.5  	          & -5.0               & 18.2                   & 45.51$^{+0.44}_{-0.44}$ \\
($-2,-2$)                         &  4.76           & -2.4	        & 2.1$^{+0.50}_{-0.90}$			  & 20.5		   & -3.2		& 2.9$^{+0.50}_{-0.90}$      	         & 20.1                   & -2.5  	          & -5.0               & 18.2                   & 45.10$^{+0.32}_{-0.35}$ \\
($-2,-4$)                         &  7.65           & -2.4	        & 2.1$^{+0.50}_{-0.90}$			  & 20.2		   & -3.2		& 2.9$^{+0.50}_{-0.90}$      	         & 20.1                   & -2.5 	          & -5.0               & 18.2                   & 45.21$^{+0.21}_{-0.24}$ \\      

\enddata
%\normalsize
\begin{footnotesize}
\tablenotetext{a}{$\sigma_{r}$ = $\pm$ 1.70 kpc, $\sigma_{U}$ = $\pm$ 0.20, $\sigma_{N_{H}}$ = $\pm$ 0.20 log(cm$^{-2}$), $\sigma_{n^{*}}$ = $\pm$ 0.20 log(cm$^{-3}$), $\sigma_{\alpha}$ = $\pm$ 0.50, $\sigma_{r^{*}}$ = $\pm$ 0.20 log(cm)} 
%\tablenotetext{b}{$\sigma_{U}$ = $\pm$ 0.20}
%\tablenotetext{c}{$\sigma_{N_{H}}$ = $\pm$ 0.20 log(cm$^{-2}$)}
%\tablenotetext{d}{$\sigma_{n^{*}}$ = $\pm$ 0.20 log(cm$^{-3}$)}
%\tablenotetext{e}{$\sigma_{\alpha}$ = $\pm$ 0.50}
%\tablenotetext{f}{$\sigma_{r^{*}}$ = $\pm$ 0.20 log(cm)}
\end{footnotesize}
\end{deluxetable}
\clearpage

%%%%%%%%%%%%%%%%%%CAPTIONS%%%%%%%%%%%%%%%%%
\clearpage

\figcaption[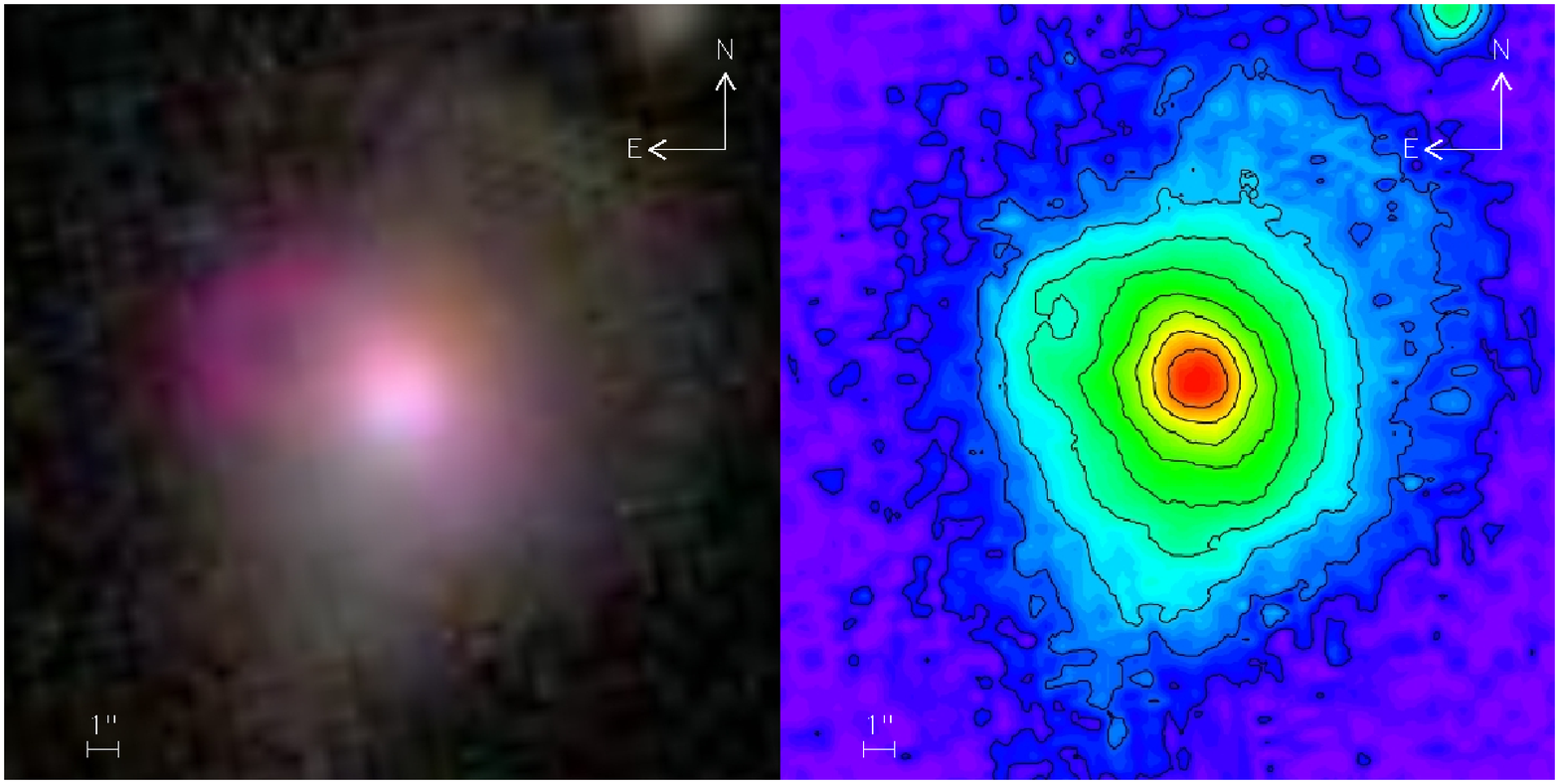]{Original Sloan optical image on the
  left, showing ionized handle of gas in the north$-$east region of
  the galaxy in purple. Right is the same SDSS image with contours
  overplotted showing the isophotal gradient of the combined g,r and i
  SDSS filters.}

\figcaption[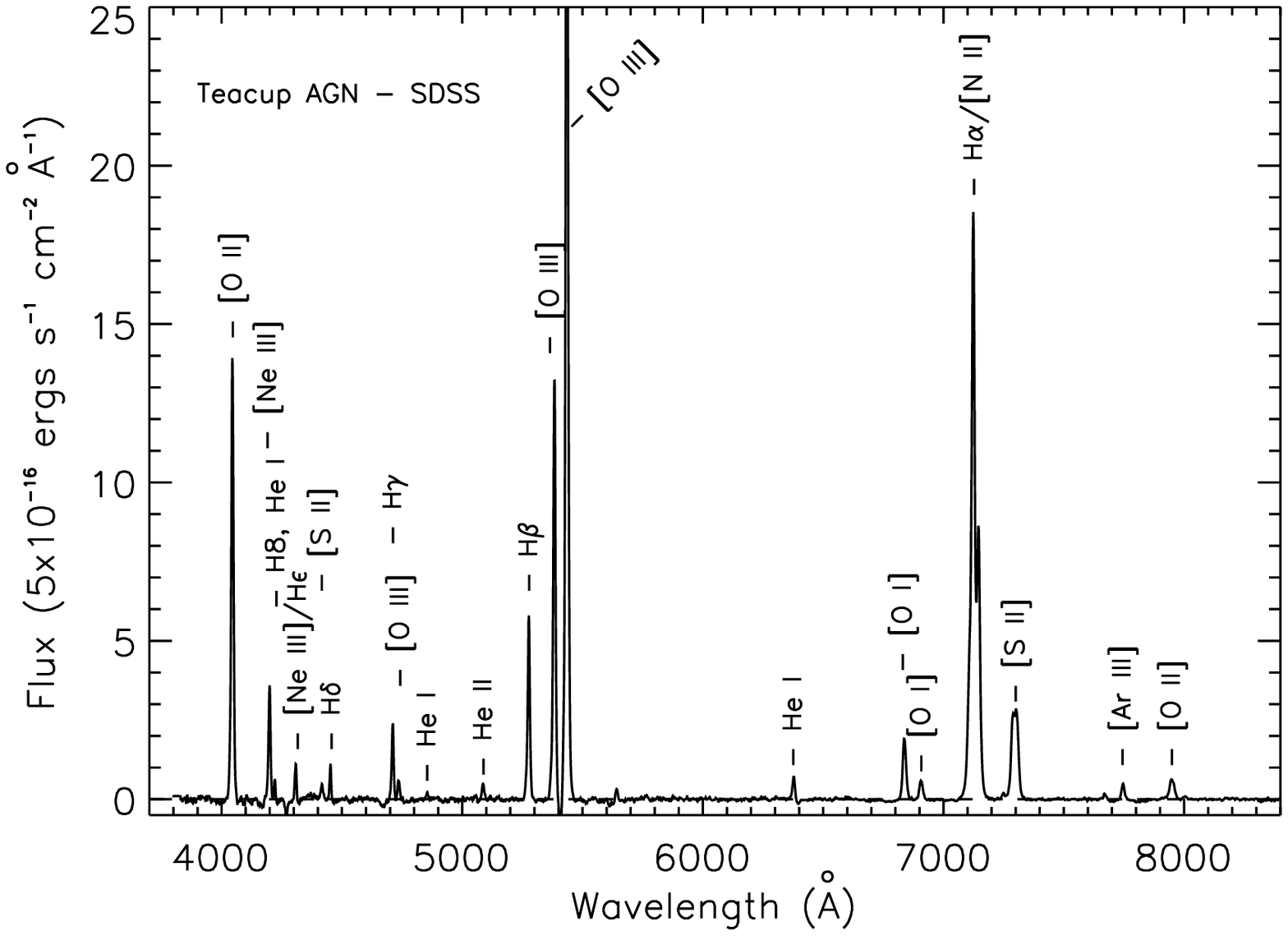]{Original SDSS spectrum with prominent lines
  identified.}

\figcaption[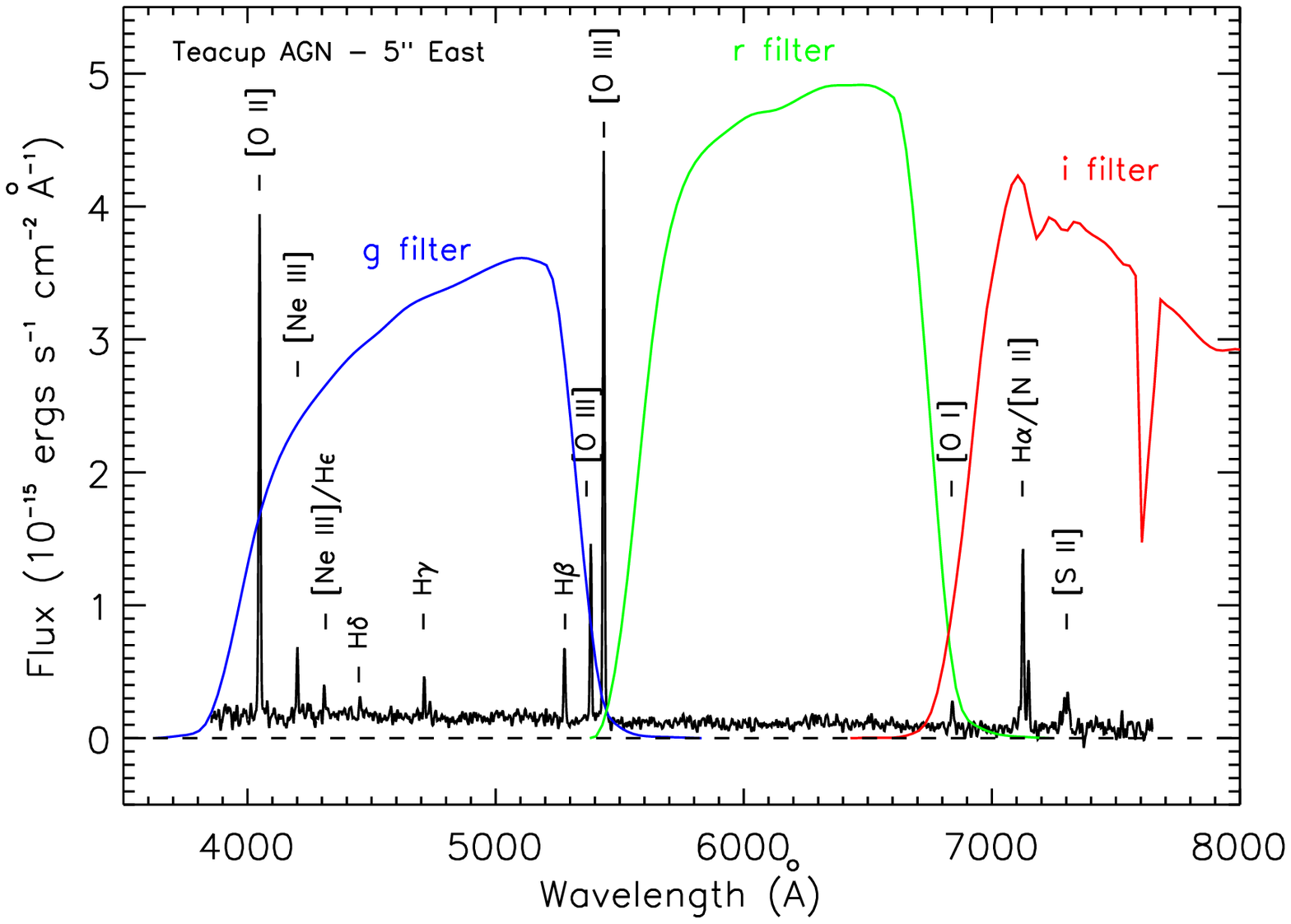]{Spectrum taken from Lowell Observatory at
  5$''$ east from the nucleus. Over plotted are the g, r and i SDSS
  filters. Note [O~III] $\lambda$ 5007 falls directly between the g
  and r bandpasses making the Teacup handle appear purple.}

\figcaption[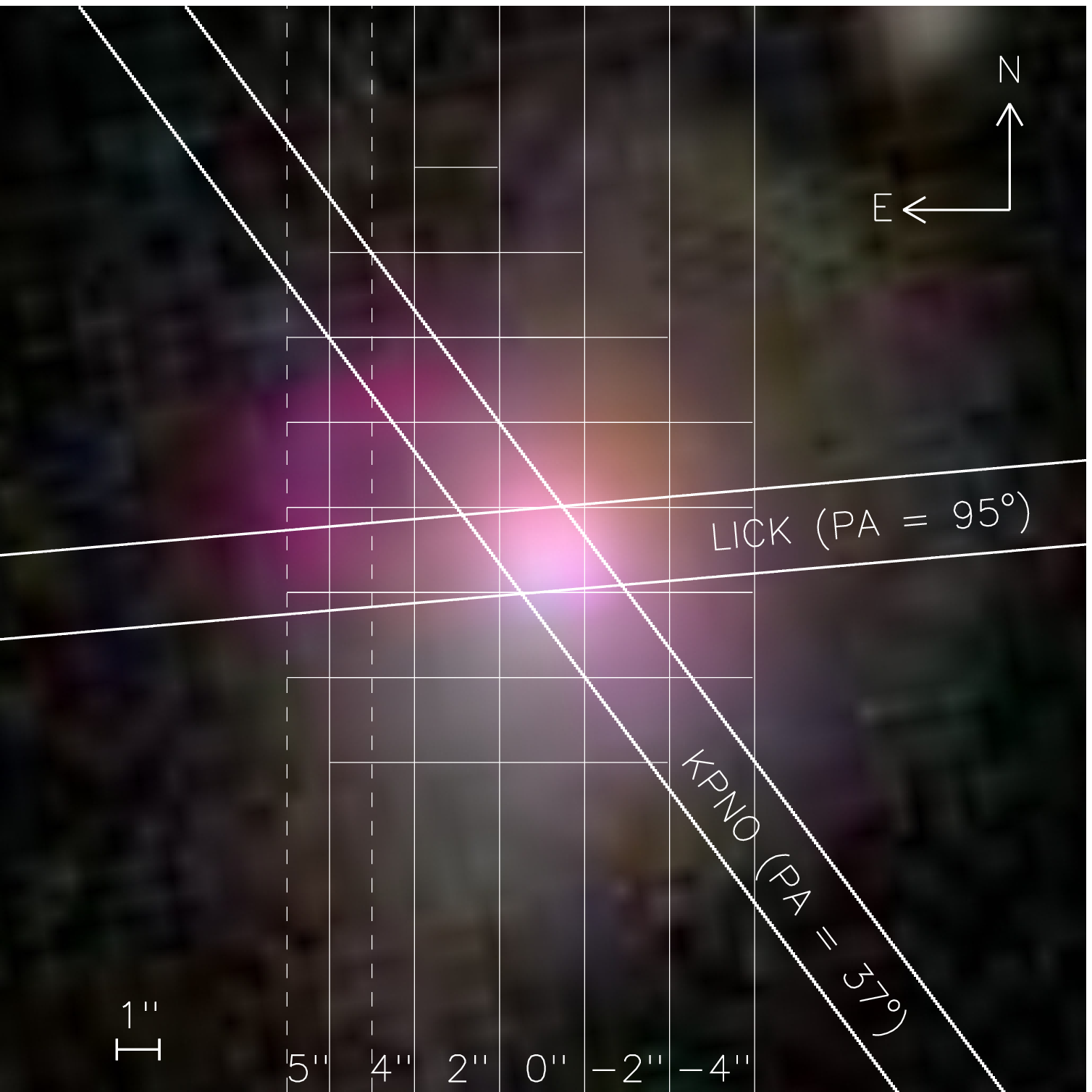]{SDSS optical image including slit positions
  for each observation. The Kast spectrograph position angle can be
  seen at 95\deg \, relative to the north - south axis of the galaxy,
  and KPNO's GoldCam spectrometer positioned at 37\deg. The vertical
  lines represent the different offset positions taken with the
  Perkins 1.8 m telescope and DeVeny spectrograph. Each extracted bin
  in the cross-dispersion direction is shown by the horizontal lines
  crossing each offset position.}

\figcaption[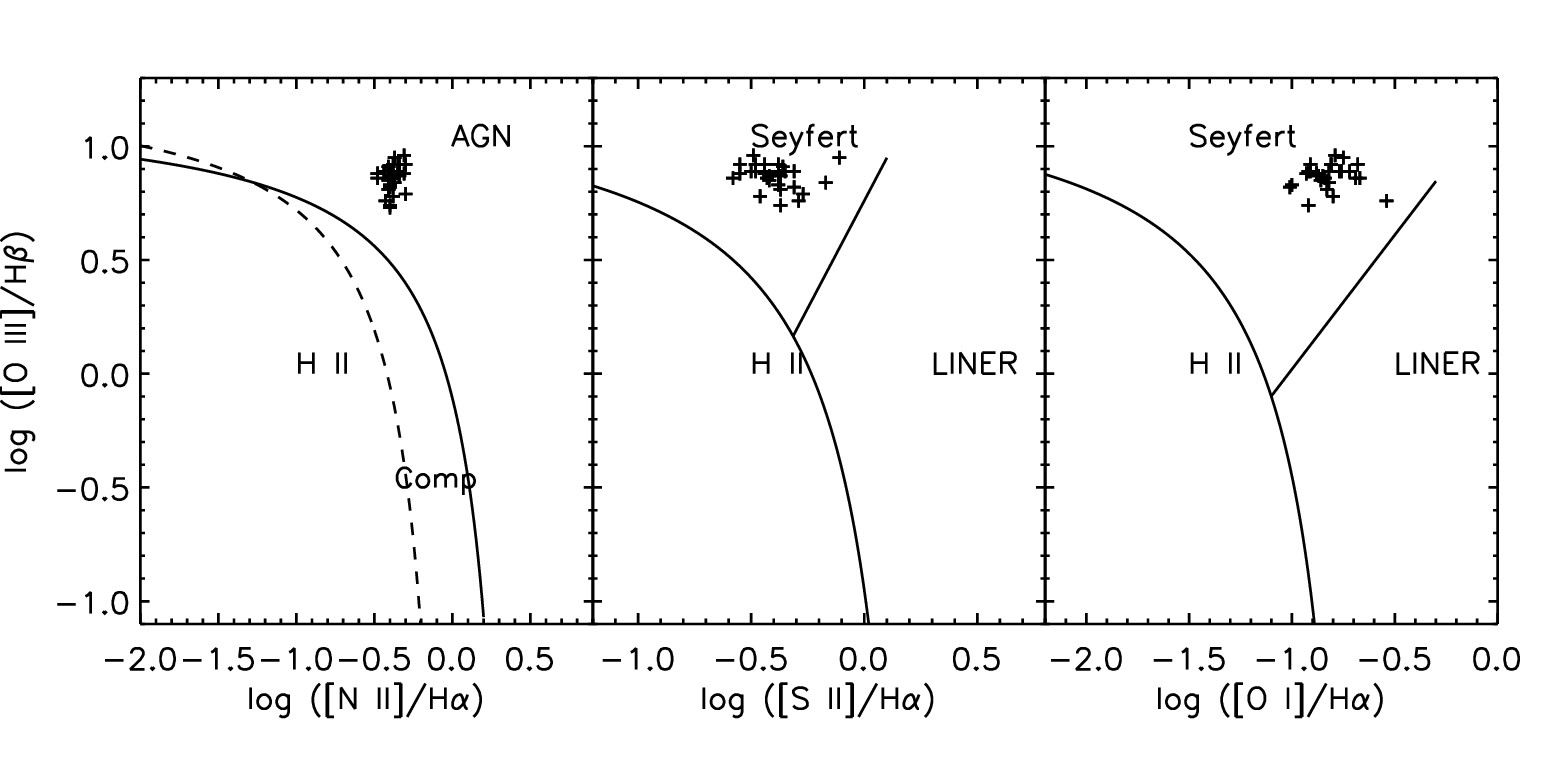]{BPT/Kewley diagrams show the Teacup is a type 2
  photoionized AGN.}

\figcaption[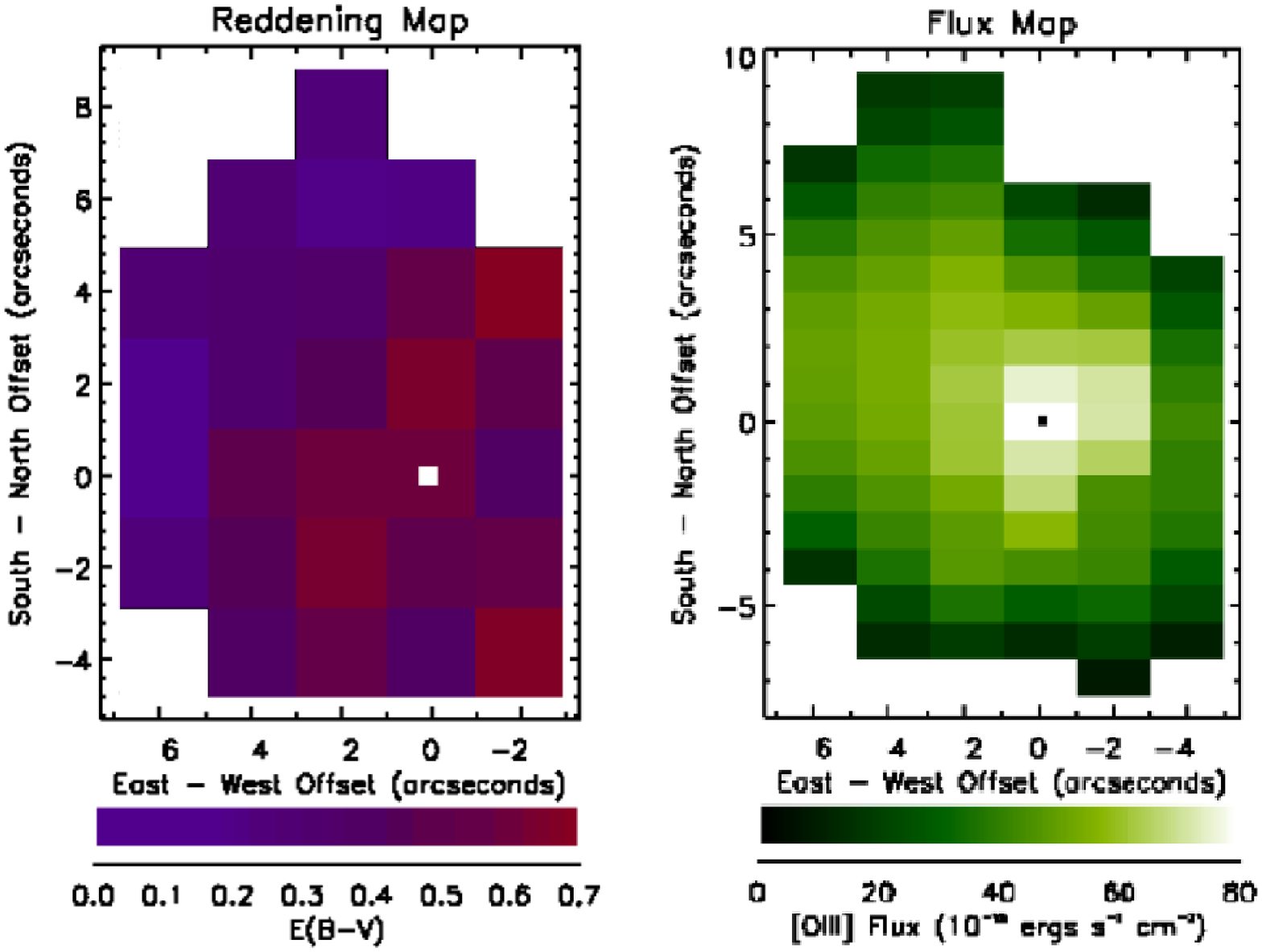]{Reddening map (left), binned in
  2$''$$\times$2$''$ squares and shows the E(B$-$V) for each given
  region. Right shows a flux map of the observed [O~III] $\lambda$
  5007 flux in 2$''$$\times$1$''$ bins.}

\figcaption[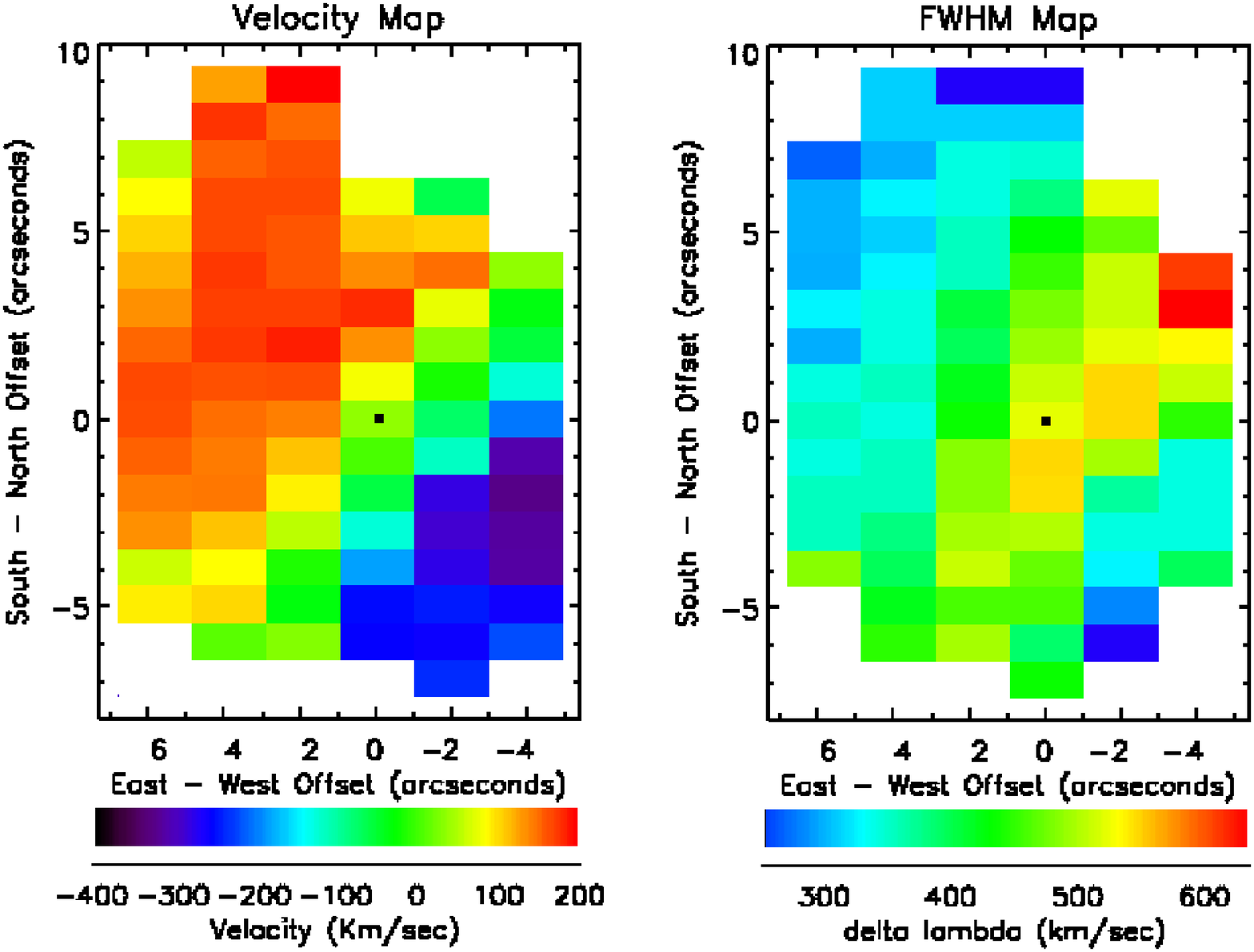]{Velocity map (left) shows the velocity of
  2$''$$\times$1$''$ bins, relative to the nucleus (marked with a
  black dot). The map on the right shows measured FWHM values in km
  s$^{-1}$ in 2$''$$\times$1$''$ bins. In this and subsequent figures,
  the 5$''$ E position has been offset by an additional 1$''$ E for
  display purposes}

\figcaption[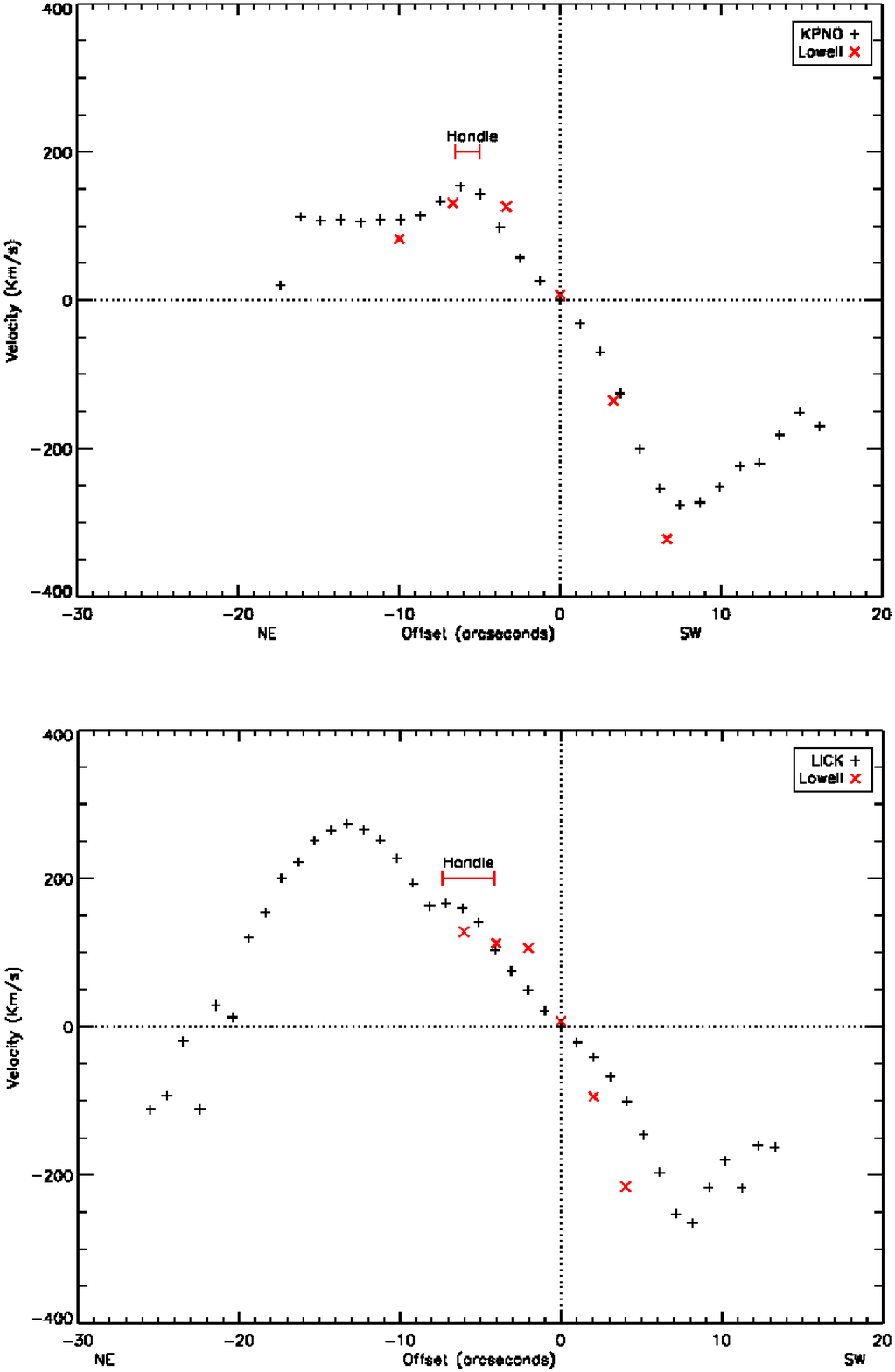]{Radial velocities measured from KPNO's
  2.1$-$m telescope at 37\deg\, and LICK's 3$-$m Shane telescope at
  95\deg\, positions angles relative to the north$-$south
  axis. Overplotted are velocities extracted from the Lowell data.}

\figcaption[Rad-Map.eps]{Left: 20cm Radio FIRST image deconvolved into
  two Gaussians. Right: SDSS g$-$band image.}

\figcaption[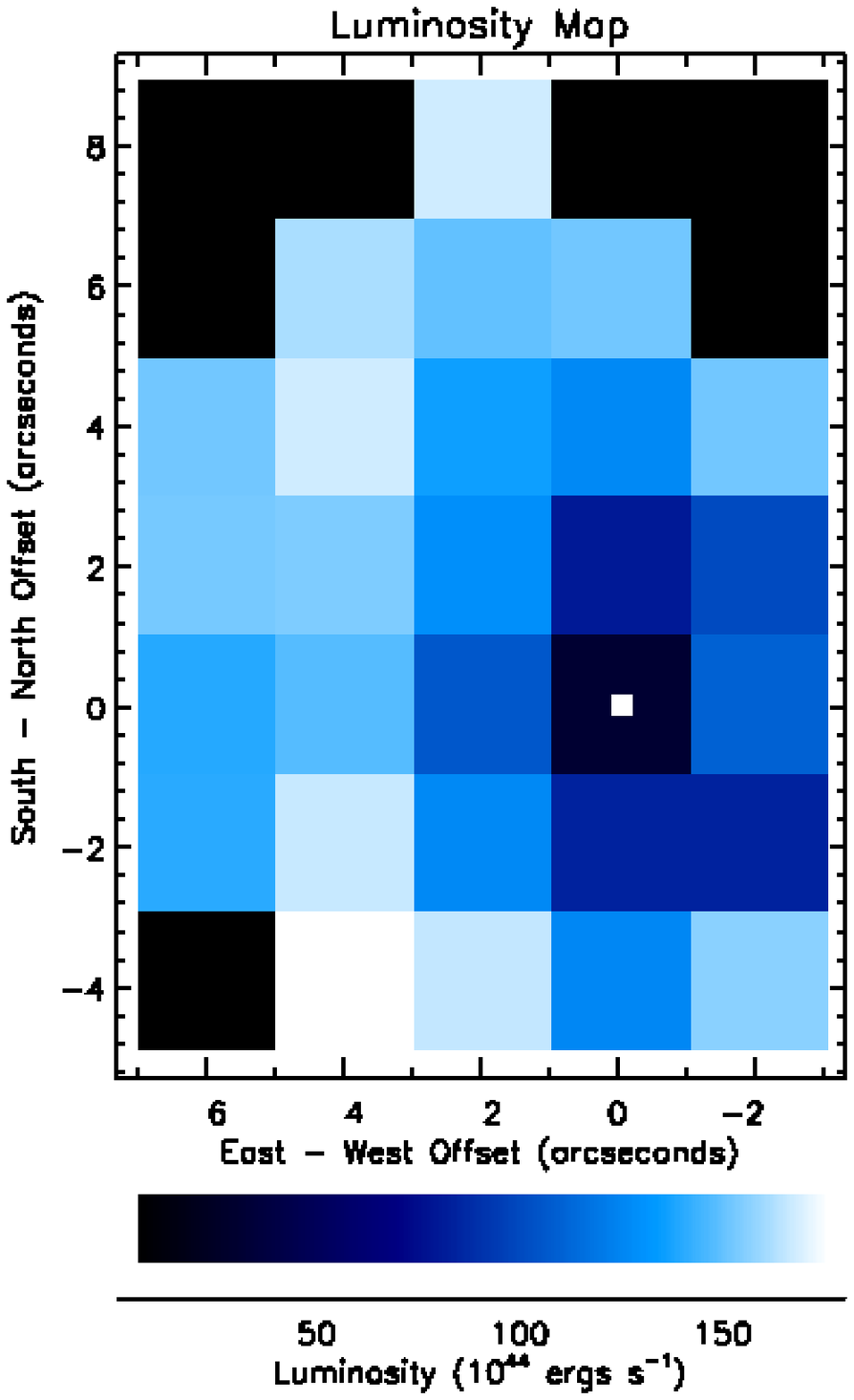]{Total luminosity per 2$''$$\times$2$''$ bin
  calculated from the ionization parameter U and electron density
  $n_{H}$. The small white box gives the location of the nucleus,
  which has a very low luminosity. The black boxes on the edges
  contain no data.}

\figcaption[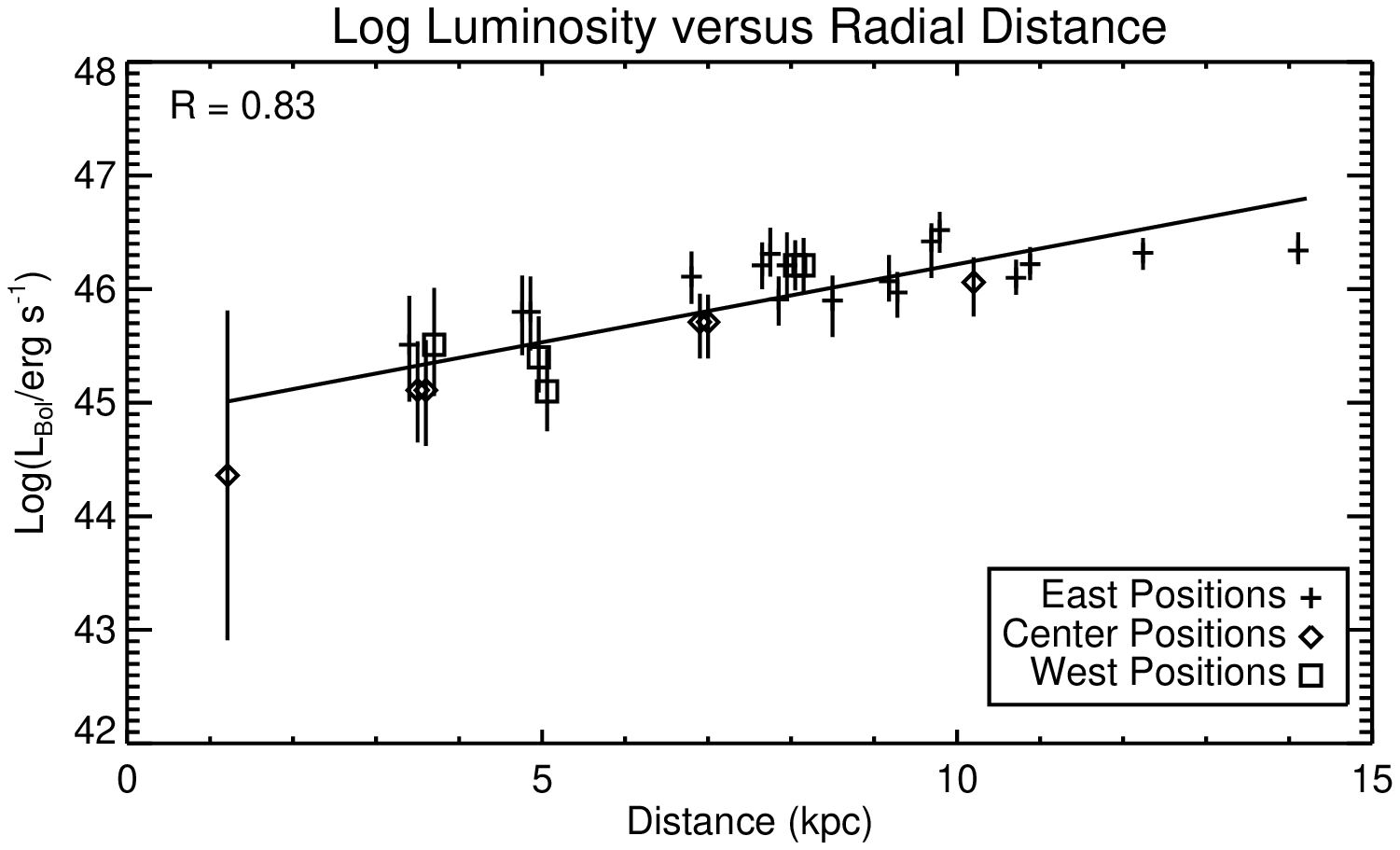]{Log of bolometric luminosity for each
  extracted position broken up by east, west and center slit
  positions, plotted versus relative distance from the nuclear
  center. Note, for ease of viewing, positions of same relative radius
  have been slightly offset. A linear fit and correlation coefficient
  are shown.}

%%%%%%%%%%%%%%%%%%%FIGURES%%%%%%%%%%%%%%%%%%
\clearpage

\begin{figure}
\includegraphics[scale=0.75,angle=270]{tcup-image-panel.eps}
\\Fig.~1.
%\label{sdss raw and contour}
\end{figure}

\begin{figure}
\plotone{SDSS-Spec.eps}
\\Fig.~2.
%\label{SDSS-spec}
\end{figure}

\begin{figure}
\plotone{5E-Spec.eps}
\\Fig.~3.
%\label{5E-spec}
\end{figure}

\begin{figure}
\plotone{All-Slits.eps}
\\Fig.~4.
%\label{all-slits}
\end{figure}

\begin{figure}
\plotone{BPT.eps}
\\Fig.~5.
%\label{BPT}
\end{figure}

\begin{figure}
\includegraphics[scale=0.8,angle=90]{Red-Flux.eps}
\\Fig.~6.
%\label{red and flux maps}
\end{figure}

\begin{figure}
\includegraphics[scale=0.55,angle=90]{V-Map.eps}
\\Fig.~7.
%\label{vel and fwhm maps}
\end{figure}

\begin{figure}
\epsscale{0.9}
\plotone{Vr-Curves.eps}
\\Fig.~8.
%\label{vel-curves}
\end{figure}

\begin{figure}
\plotone{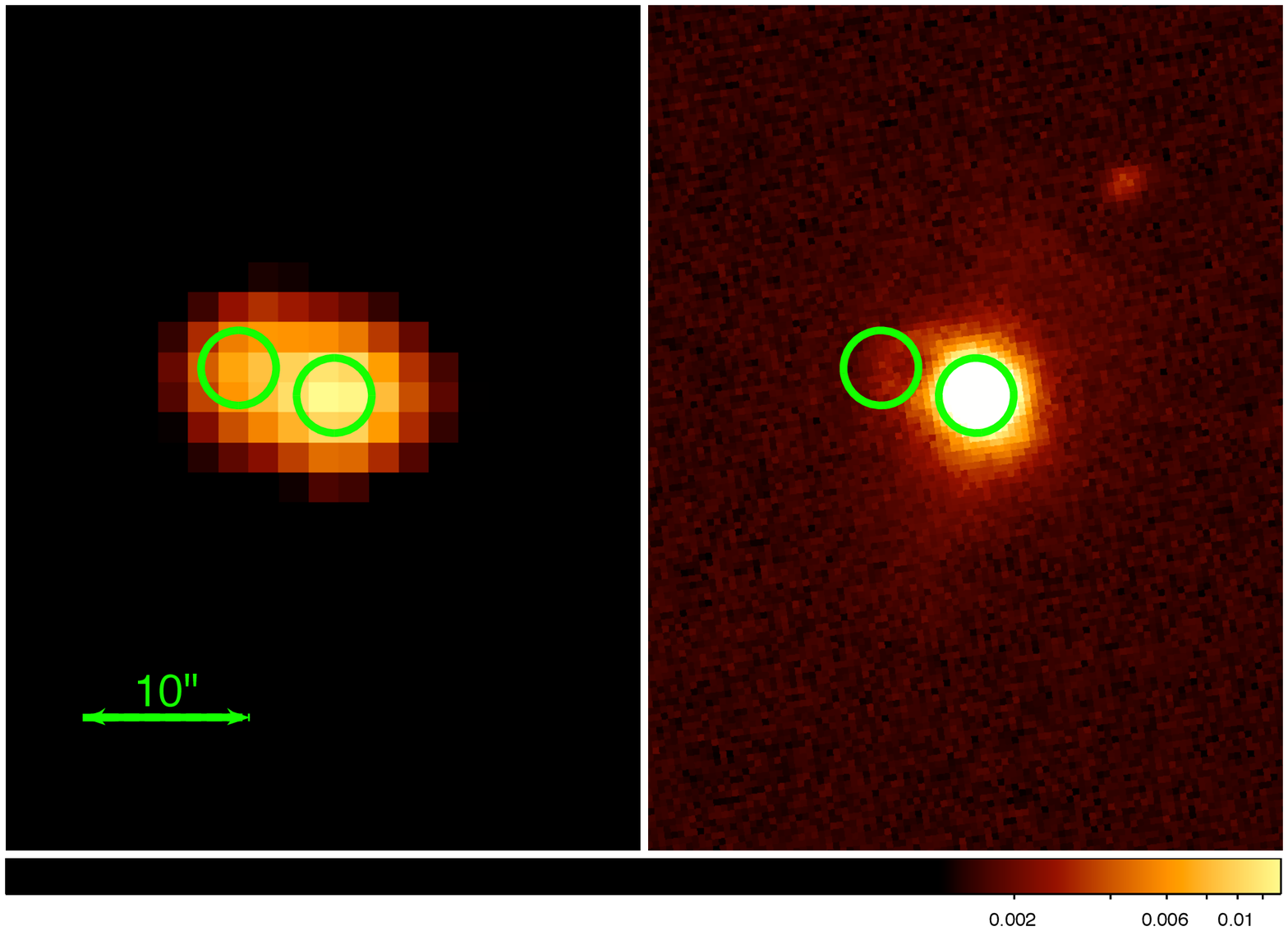}
\\Fig. ~9.
%\label{radio maps}
\end{figure}

\begin{figure}
\epsscale{0.85}
\plotone{L-Map.eps}
\\Fig.~10.
%\label{lum-map}
\end{figure}

%\begin{figure}
%\includegraphics[scale=1.0,angle=0]{fig9.eps}
%\\Fig.~9.
%\end{figure}

\begin{figure}
\epsscale{1.0}
\plotone{L-vs-R.eps}
\\Fig.~11.
%\label{lum v r}
\end{figure}

\end{document}